\documentclass[11pt,a4paper]{emulateapj}
\usepackage{epsfig}
\usepackage{natbib}
\usepackage{graphicx}

\begin{document}

\title{CANDELS Multiwavelength Catalogs: Source Identification and
Photometry in the CANDELS UKIDSS Ultra-deep Survey Field}

\author{
Audrey Galametz\altaffilmark{1},  
Andrea Grazian\altaffilmark{1}, 
Adriano Fontana\altaffilmark{1},
Henry C. Ferguson\altaffilmark{2},
M. L. N. Ashby\altaffilmark{3},
Guillermo Barro\altaffilmark{4},
Marco Castellano\altaffilmark{1},
Tomas Dahlen\altaffilmark{2},
Jennifer L. Donley\altaffilmark{5},
Sandy M. Faber\altaffilmark{4},
Norman Grogin\altaffilmark{2}, 
Yicheng Guo\altaffilmark{4,6},
Kuang-Han Huang\altaffilmark{2,7},
Dale D. Kocevski\altaffilmark{8},
Anton M. Koekemoer\altaffilmark{2}, 
Kyoung-Soo Lee\altaffilmark{9},
Elizabeth J. McGrath\altaffilmark{10},
Michael Peth\altaffilmark{7},
S. P. Willner\altaffilmark{3},
Omar Almaini\altaffilmark{11},
Michael Cooper\altaffilmark{12}, 
Asantha Roshan Cooray\altaffilmark{12},
Christopher J. Conselice\altaffilmark{11},
Mark Dickinson\altaffilmark{13},
James S. Dunlop\altaffilmark{14},
G. G. Fazio\altaffilmark{3},
Sebastien Foucaud\altaffilmark{15},
Jonathan P. Gardner\altaffilmark{16}
Mauro Giavalisco\altaffilmark{6},
N. P. Hathi\altaffilmark{17},
Will G. Hartley\altaffilmark{11},
David C. Koo\altaffilmark{4},
Kamson Lai\altaffilmark{2},
Duilia F. de Mello\altaffilmark{18},
Ross J. McLure\altaffilmark{14},
Ray A. Lucas\altaffilmark{2},
Diego Paris\altaffilmark{1},
Laura Pentericci\altaffilmark{1},
Paola Santini\altaffilmark{1},
Chris Simpson\altaffilmark{19},
Veronica Sommariva\altaffilmark{1},
Thomas Targett\altaffilmark{14},
Benjamin J. Weiner\altaffilmark{20},
Stijn Wuyts\altaffilmark{21}
and the CANDELS team}

\altaffiltext{1}{INAF - Osservatorio di Roma, Via Frascati 33, I-00040, Monteporzio, Italy [e-mail: {\tt audrey.galametz@oa-roma.inaf.it}]}
\altaffiltext{2}{Space Telescope Science Institute, Baltimore, MD, USA}
\altaffiltext{3}{Harvard-Smithsonian Center for Astrophysics, Cambridge, MA, USA}
\altaffiltext{4}{UCO/Lick Observatory, Department of Astronomy and Astrophysics, University of California, Santa Cruz, CA, USA}
\altaffiltext{5}{Los Alamos National Laboratory, Los Alamos, NM, USA}
\altaffiltext{6}{Department of Astronomy, University of Massachusetts, Amherst, USA}
\altaffiltext{7}{Department of Physics and Astronomy, The Johns Hopkins University, Baltimore, MD, USA}
\altaffiltext{8}{Department of Physics and Astronomy, University of Kentucky, Lexington, USA}
\altaffiltext{9}{Department of Physics, Purdue University, West Lafayette, USA}
\altaffiltext{10}{Department of Physics and Astronomy, Colby College, Waterville, USA}
\altaffiltext{11}{The School of Physics and Astronomy, University of Nottingham, Nottingham, UK}
\altaffiltext{12}{Department of Physics and Astronomy, University of California, Irvine, USA}
\altaffiltext{13}{National Optical Astronomy Observatories, Tucson, AZ, USA}
\altaffiltext{14}{Institute for Astronomy, University of Edinburgh, Royal Observatory, Edinburgh, UK}
\altaffiltext{15}{National Taiwan Normal University, Taiwan, Republic of China}
\altaffiltext{16}{NASA Goddard Space Flight Center, Astrophysics Science Division, Observational Cosmology Laboratory, Greenbelt, MA, USA}
\altaffiltext{17}{Observatories of the Carnegie Institution for Science, Pasadena, CA, USA}
\altaffiltext{18}{Physics Department, The Catholic University of America, Washington, DC, USA}
\altaffiltext{19}{Astrophysics Research Institute, Liverpool John Moores University, Birkenhead, UK}
\altaffiltext{20}{Steward Observatory, University of Arizona, Tucson, AZ 85721}
\altaffiltext{21}{Max-Planck-Institut fur extraterrestrische Physik, Garching bei Munchen, Germany}

\begin{abstract}
We present the multiwavelength --- ultraviolet to mid-infrared --- catalog of the UKIDSS Ultra-Deep Survey 
(UDS) field observed as part of the Cosmic Assembly Near-infrared Deep Extragalactic Legacy Survey (CANDELS). Based 
on publicly available data, the catalog includes: the CANDELS data from the {\it Hubble} Space Telescope (near-infrared WFC3 
$F125W$ and $F160W$ data and visible ACS $F606W$ and $F814W$ data), $u$-band data from CFHT/Megacam, $B$, $V$, 
$R_c$, $i'$ and $z'$ band data from Subaru/Suprime-Cam, $Y$ and $K_s$ band data from VLT/HAWK-I, $J$, $H$ and 
$K$ bands data from UKIDSS (Data Release 8), and {\it Spitzer}/IRAC data ($3.6$, $4.5$ from SEDS, 
$5.8$ and $8.0\mu$m from SpUDS). The present catalog is $F160W$-selected and contains $35932$ sources 
over an area of $201.7$ square arcmin and includes radio and X-ray detected sources and spectroscopic 
redshifts available for $210$ sources. 

\end{abstract}
\keywords{galaxies: photometry Ð methods: data analysis Ð techniques: image processing}

\section{Introduction}

The Cosmic Assembly Near-infrared Deep Extragalactic Legacy Survey \citep[][CANDELS; PIs S. Faber, 
H. Ferguson]{Grogin2011, Koekemoer2011}, a 902-orbit Multi-Cycle Treasury (MCT) program, is the largest 
project ever approved for the {\it Hubble} Space Telescope (HST). CANDELS is currently obtaining {\it HST} 
observations of five well-studied sky regions: the GOODS-North and GOODS-South fields (Giavalisco et al. 2004) as 
well as subsections of the Extended Groth Strip (EGS; Davis et al. 2007), COSMOS 
(Scoville et al. 2007), and the UKIDSS Ultra-Deep Survey (UDS; Lawrence et al. 2007; Cirasuolo et al. 2007).
Most of the observations make use of the Wide Field Camera 3 (WFC3)/IR as prime instrument and the Advanced 
Camera for Surveys (ACS) in parallel. These five CANDELS fields were natural choices because of the 
large number of ancillary data available in these fields. In particular, they were all covered by deep {\it Spitzer}/IRAC 
$3.6\mu$m and $4.5\mu$m imaging within the {\it Spitzer} Extended Deep Survey (SEDS; PI: G. Fazio; Ashby et 
al.~ resubmitted).

The CANDELS data are made public right after acquisition. Because of the treasury aspect of the CANDELS
project, the team will provide, alongside the final reduced HST mosaics, multiwavelength catalogs for each of the 
five CANDELS fields. We present in this paper the efforts of the CANDELS Multiwavelength Group to converge
to a unique and homogeneous recipe to build all CANDELS legacy multiwavelength catalogs. These catalogs will include 
the best (higher resolution and deepest) available ultraviolet to mid-infrared data ever taken in each of the five CANDELS 
fields, either from ground-based or space telescopes. The CANDELS multiwavelength group efforts first 
concentrated on the two first completed fields, namely the UDS (present paper) and the GOODS-S.
We refer to Guo et al.~(submitted) for the details on the building of the multiwavelength catalog for GOODS-S. 
The present paper illustrates the catalog building methodology by describing each step of the creation of the 
first released multiwavelength catalog in the UDS field, the first CANDELS field that was fully observed by HST. 

The CANDELS UDS field (R.A. = 02:17:37.5; Dec. = -05:12:00) resides within the original UDS field, observed 
by a large range of ultraviolet to mid-infrared data (as well as in X-ray and radio), namely CFHT/Megacam 
$u$-band, Subaru/Suprime-Cam $B$, $V$, $R_c$, $i'$ and $z'$, CANDELS {\it HST}/ACS ($F606W$, $F814W$) 
and {\it HST}/WFC3 ($F125W$, $F160W$), VLT/HAWK-I ($Y$, $K_s$), UKIRT/WFCAM ($J$, $H$, $K$) and 
{\it Spitzer}/IRAC ($3.6$, $4.5$, $5.8$ and $8.0\mu$m).

Multiwavelength imaging provides a great insight into the properties of astronomical objects. Sources can appear
luminous at some wavelength and fade completely at others since different filters intrinsically reveal different 
properties of the same object. Another technical difficulty lies in analyzing simultaneously the inhomogeneous 
multiwavelength dataset at hand, often produced by instruments with different characteristics (e.g.,~ground- versus 
space-based telescopes). For example, isolated sources in high-resolution data such as HST can become 
rapidly blended with their closer neighbors in low-resolution data. This thus prevents a reliable estimation of the 
source photometry. The main advantage of the CANDELS dataset is precisely the existence for each field of 
high-resolution deep images that provide {\it a-priori} information on a source (position, morphology etc.) that help 
derive photometry of their counterparts in lower-resolution data. The present CANDELS UDS catalog --- a 
HST/WFC3 $F160W$-detected source catalog --- represents a major improvement over past individual catalogues
in this field (although the present catalog concentrated on the CANDELS field of view only). Using the positional information 
from the CANDELS HST data, it is possible to estimate fluxes and upper limits reliably to much fainter limits 
in the non-HST data, and to formally incorporate the covariance due to blended sources into the flux uncertainties.

The present paper is organized as follows. Section 2 gives a thorough description of the data available in the 
CANDELS UDS field. Section 3 presents the source extraction on the HST/WFC3 $F160W$ image, in particular 
the adopted two-step `cold + hot' extraction technique. Photometry of the HST data is described in Section 4 and
of the lower-resolution data (i.e.,~ground-based and {\it Spitzer}/IRAC) in Section 5. Section 5 deals, in particular, 
with the preliminary steps required to prepare the low-resolution data for the use of the adopted Template-fitting photometry 
software, TFIT. Section 6 presents the final multiwavelength catalog. Validation tests on the catalog 
photometry are listed in Section 7, and a summary is presented in Section 8.

All magnitudes are given in the AB photometric system, and we adopt a $\Lambda$CDM cosmology 
with $H_0 = 70$ km s$^{-1}$ Mpc$^{-1}$, $\Omega_m = 0.3$ and $\Omega_{\Lambda} = 0.7$

The CANDELS UDS multiwavelength catalog and its associated documentation --- as well as the total 
system throughput curves of all the filters included in the catalog --- are made publicly available on 
the CANDELS website\footnote[1]{http://candels.ucolick.org} and on the Mikulski Archive for Space 
Telescopes (MAST)\footnote[2]{http://archive.stsci.edu/}. The catalog will also be made available via 
the on-line version of the article, the Centre de Donn\'{e}es astronomiques de Strasbourg (CDS) 
as well as in the Rainbow Database\footnote[3]{Europe: https://rainbowx.fis.ucm.es/Rainbow\_navigator\_public/. 
US: https://arcoiris.ucolick.org/Rainbow\_navigator\_public/. The web-interface 
to the Rainbow Database features a query menu that allows the user to search for individual galaxies, create 
subsets of the complete sample based on different filters, or inspect cutouts of the galaxies in any of the
available bands. It also includes a cross-matching tool to compare against user uploaded catalogs.} 
\citep{Perez2008, Barro2011}.

\section{Data}

The (normalized) total system throughput curves and fields of view of all the data included in the CANDELS UDS 
multiwavelength catalog are shown in Figures~\ref{trans} and \ref{area} respectively. Table~\ref{data} 
summarizes the data. It also provides, for each band, a limiting magnitude estimate (without aperture correction) 
derived from the rms in an aperture of one full width at half maximum (FWHM) radius, at a $5\sigma$ level.

\subsection{The CANDELS HST data}

The CANDELS UDS field was covered by {\it HST} using a mosaic grid of tiles that was observed
over two epochs. During each epoch, the field was imaged in one orbit ($\sim2000$s) with the Wide 
Field Camera 3 (WFC3) split into $F125W$ ($1/3$ orbit) and $F160W$ ($2/3$ orbit) together 
with parallel exposures using the Advanced Camera for Surveys (ACS) in $F606W$ and $F814W$.

The WFC3 mosaics are composed of a grid of $4 \times 11$ tiles (see Fig.~3; top) at spacing 
intervals that maximize the coverage without introducing gaps between tiles, resulting in a final 
rectangular field of view of $\sim22.3\arcmin \times 9\arcmin$. The long axis is at a position angle 
of $-90$~degrees. Exposures were oriented so that the ACS parallels are offset
along the long axis of the mosaic, producing a similar-sized mosaic overlapping most of the 
WFC3 mosaic, except at its edges where some tiles are only covered by WFC3 or by ACS.
We refer to \citet[][Figure~14]{Grogin2011} and \citet[][Figures~17 to 22]{Koekemoer2011} for details 
on the {\it HST} data set, mosaics and data reduction. The final UDS {\it HST} mosaics (e.g.,~drizzled science 
images and inverse variance weight images) are publicly accessible via the STScI 
archive\footnote[4]{\tt http://archive.stsci.edu/prepds/candels/}.

\subsection{Ground-based Imaging}

The CANDELS UDS field is covered by a large number of ground-based data. Deep near-infrared data 
of an area of $0.8$ square degrees (including the CANDELS field) were obtained in $J$, 
$H$ and $K$ as part of the UKIRT Infrared Deep Sky Survey \citep[UKIDSS;][]{Lawrence2007} using the 
Wide Field Camera 
(WFCAM) on the UKIRT telescope. The current public data release (UKIDSS DR8) reaches median 
depths of  $J=24.9$, $H=24.2$, $K=24.6$ ($5\sigma$). 
Intermediate data releases including images are available from the WFCAM Science 
Archive\footnote[5]{\tt http://surveys.roe.ac.uk/wsa/}. Almaini et al.~(in prep.) provide details on the data.

The full UKIDSS field of view was also imaged by the Canada France Hawaii Telescope (CFHT) 
in the $u$-band with MegaCam (PIs: O. Almaini, S. Foucaud; Almaini et al. in prep.).

The CANDELS UDS field was also observed in other near-infrared bands as part of the HAWK-I UDS and 
GOODS-S survey (HUGS; VLT large program ID 186.A-0898, PI: A. Fontana; Fontana et al.~in prep.) 
using the High Acuity Wide field K-band Imager (HAWK-I) on VLT. About $95$\% of the CANDELS UDS was covered 
by three HAWK-I pointings in both the $Y$ and $K_s$ bands. All three pointings were imaged in $Y$ and $K_s$ 
for $\sim8$~hours and $\sim13$~hours respectively.
We refer to Fontana et al.~in prep. for more details on the HUGS data. 

A large set of optical imaging observations in the UDS field were taken with Suprime-Cam 
on the Subaru Telescope as part of the Subaru/XMM-Newton Deep Survey (SXDS) in $B$, $V$, $R_c$, 
$i'$ and $z'$-band. These data reach a $3\sigma$ limit ($2$-arcsec diameter aperture) magnitude of
$B = 28.4$, $V = 27.8$, $R = 27.7$, $i = 27.7$ and $z = 26.6$ \citep{Furusawa2008}. 
Mosaics (registered to the CANDELS astrometry) are available 
online\footnote[6]{\tt http://www.roe.ac.uk/$\sim$ciras/Scientific\_Research.html} \citep[see also][]{Cirasuolo2010}. 

\begin{table*}
\caption{UDS dataset}
\label{data}
\centering
\begin{tabular}{l l c c c l}
\hline
Instrument	&	Filter 	&	Central		&	FWHM		&	Limiting 		&	Survey$^{\mathrm{a}}$		\\
			&			&	wavelength	&			&	magnitude	&						\\
			&			&	(nm)			&	(arcsec)	&	($5\sigma$, 1 FWHM radius, AB)	&		\\      
\hline 
CFHT/MegaCam			&	$u$			&	$386$	&	$0.86$	&	$27.68$	&	(1)	\\	
Subaru/Suprime-Cam		&	$B$			&	$450$	&	$0.82$	&	$28.38$	&	(2)	\\	
						&	$V$			&	$548$	&	$0.82$	&	$28.01$	&	(2)	\\	
						&	$R_c$		&	$650$	&	$0.80$	&	$27.78$	&	(2)	\\	
						&	$i'$			&	$768$	&	$0.82$	&	$27.69$	&	(2)	\\	
						&	$z'$			&	$889$	&	$0.81$	&	$26.67$	&	(2)	\\	
{\it HST}/ACS				&	$F606W$		&	$598$	&	$0.10$	&	$28.49$	&	(3)	\\	
						&	$F814W$		&	$791$	&	$0.10$	&	$28.53$	&	(3)	\\	
{\it HST}/WFC3				&	$F125W$		&	$1250$	&	$0.20$	&	$27.35$	&	(3)	\\	
						&	$F160W$		&	$1539$	&	$0.20$	&	$27.45$	&	(3)	\\	
VLT/HAWK-I$^{\mathrm{b}}$	&	$Y$	 		&	$1019$	&	$0.42/0.52/0.49$	&	$27.05/26.73/26.69$	&	(4)	\\	
						&	$K_s$		&	$2147$	&	$0.36/0.42/0.37$	&	$26.16/25.92/25.98$	&	(4)	\\	
UKIRT/WFCAM				&	$J$			&	$1251$	&	$0.76$	&	$25.63$	&	(1)	\\	
						&	$H$			&	$1636$	&	$0.80$	&	$24.76$	&	(1)	\\	
						&	$K$			&	$2206$	&	$0.70$	&	$25.39$	&	(1)	\\	
{\it Spitzer}/IRAC			&	$3.6\mu$m	&	$3562$	&	$\sim1.9$	&	$24.72$	&	(5)	\\	
						&	$4.5\mu$m	&	$4512$	&	$\sim1.9$	&	$24.61$	&	(5)	\\	
						&	$5.8\mu$m	&	$5686$	&	$2.08$	&	$22.30$	&	(6)	\\	
						&	$8.0\mu$m	&	$7936$	&	$2.20$	&	$22.26$	&	(6)	\\	
\hline     
\end{tabular}
\begin{list}{}{}
\item[$^{\mathrm{a}}$] (1) UKIDSS - Almaini et al.~in prep. (2) SXDS - Furusawa et al.~2008 (3) CANDELS - Koekemoer et al.~2011
(4) HUGS - Fontana et al.~in prep (5) SEDS - Ashby et al.~resubmitted. (6) SpUDS.
\item[$^{\mathrm{b}}$] FWHM and limiting magnitudes are provided for the three HAWK-I pointings following the scheme 
Pointing1/Pointing2/Pointing3 (i.e.,~Central/West/East; see Fontana et al.~in prep).
\end{list}
\end{table*}

\begin{figure}
\begin{center}
\includegraphics[width=8.8cm]{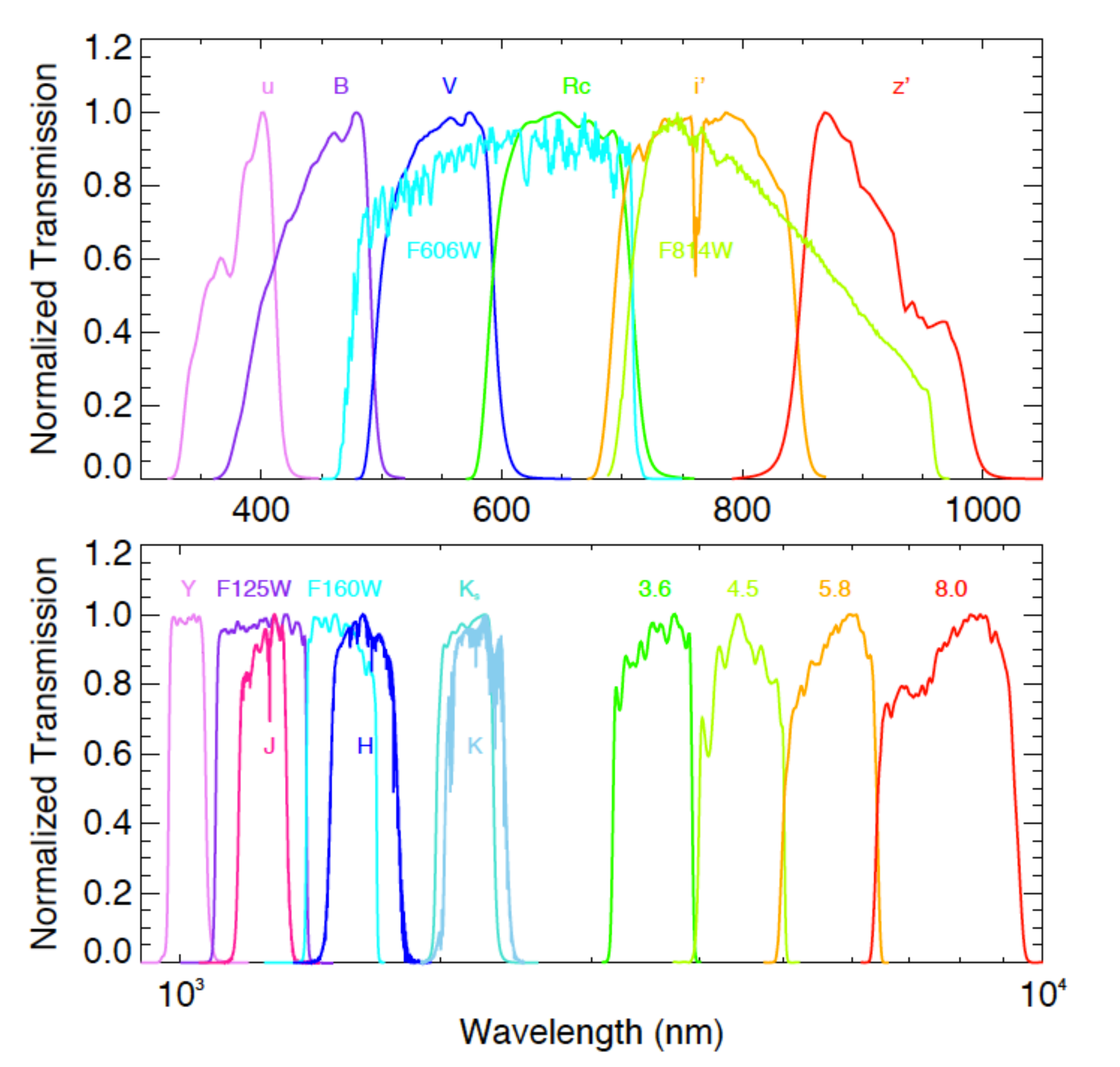}
\end{center}
\caption{(Normalized) total system throughput (quantum efficiency) of the optical (top panel), and near to mid-infrared filters 
(bottom; log scale on x axis) of data available in the UDS CANDELS field.}
\label{trans}
\end{figure}

\begin{figure}
\begin{center}
\includegraphics[width=8.8cm]{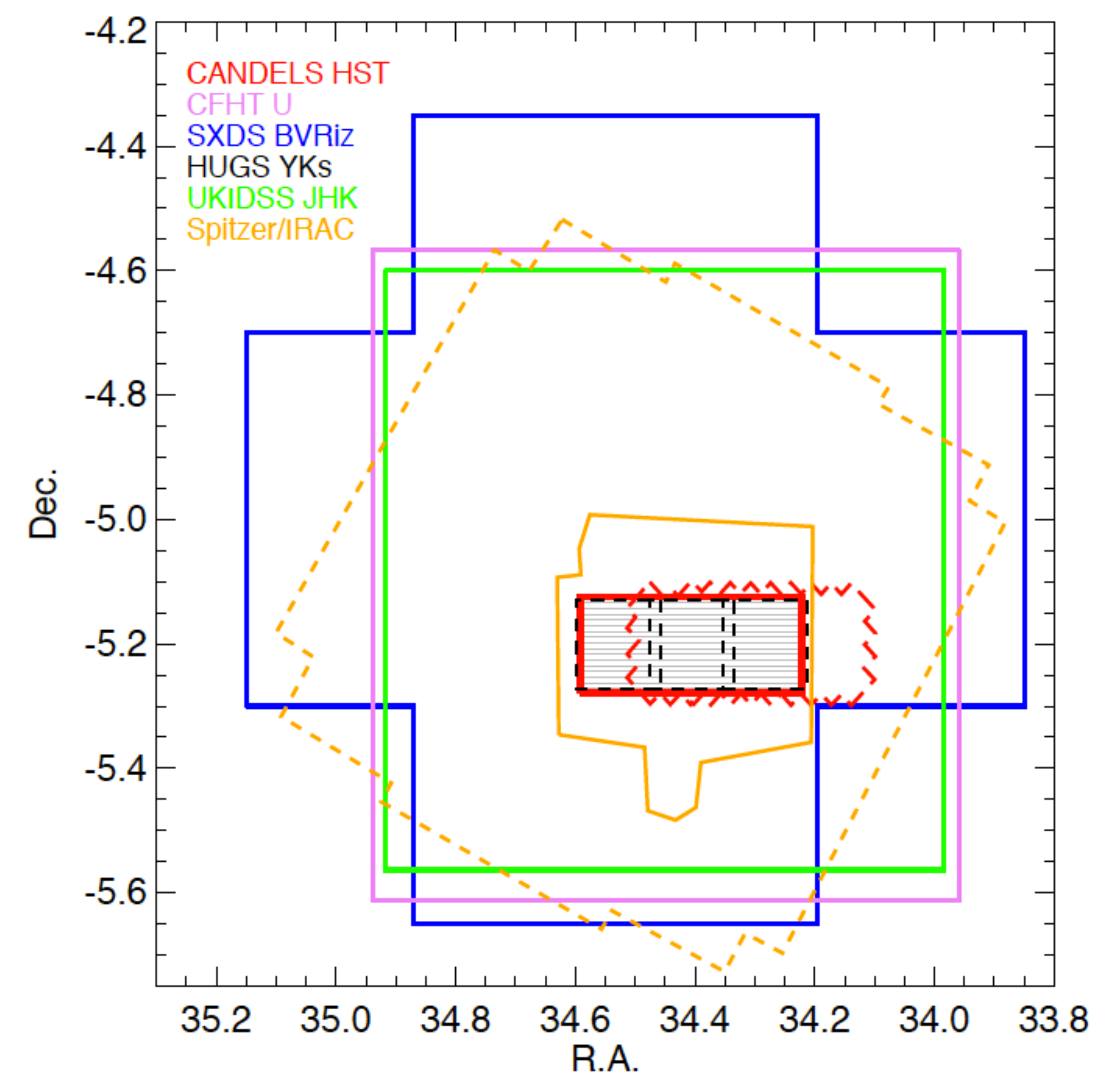}
\end{center}
\caption{Coverage of the imaging data in the UDS field: {\it HST} WFC3 (red solid and grey hatched 
region; the detection band for the present multiwavelength catalog) and ACS (red dashed), CFHT $u$-band 
(purple; Almaini et al.~in prep.), Subaru data ($BVR_ci'z'$; blue; Furusawa et al.~2008), 
UKIDSS ($JHK$; green; Almaini et al.~in prep.), HUGS ($YK_s$; black dashed; $3$ pointings; 
Fontana et al.~in prep.), {\it Spitzer}/IRAC SEDS ($3.6$ and $4.5\mu$m; orange solid; Ashby et 
al.~resubmitted) and SpUDS (four IRAC bands; orange dashed).}
\label{area}
\end{figure}

\subsection{{\it Spitzer}/IRAC data}

Apart from the shallow ($4 \times 30$s) IRAC coverage obtained by the {\it Spitzer} Wide-Area 
Infrared Extragalactic Survey \citep[SWIRE; ][]{Lonsdale2003}, the UDS was surveyed as part 
of a {\it Spitzer} cycle-4 Legacy Program: the {\it Spitzer} UKIDSS Ultra Deep Survey (SpUDS 
hereafter; PI: J. Dunlop). SpUDS covers an area of about $1$ square degree (see Figure~\ref{area}) 
in the four IRAC channels ($3.6$, $4.5$, $5.8$ and $8.0\mu$m) and reaches 
a depth of $24.7$~mag ($1\sigma$) at $4.5\mu$m. Images (and SExtractor catalogs) for SpUDS 
can be found online\footnote[7]{\tt http://irsa.ipac.caltech.edu/data/SPITZER/SpUDS/}. 

A subregion of SpUDS ($0.40$~deg$^2$; see Figure~\ref{area}) --- that also contains the full CANDELS 
UDS field --- has recently been observed as part of the {\it Spitzer} Extended Deep Survey (SEDS hereafter; 
PI: G. Fazio) during the {\it Spitzer} Warm Mission\footnote[8]{\tt http://www.cfa.harvard.edu/SEDS} 
at $3.6$ and $4.5\mu$m. A final UDS mosaic incorporating all coextensive exposures from SEDS, SWIRE 
and SpUDS was generated by the SEDS team so as to reach the desired $12$~hours total integration time/pixel 
within the SEDS footprint (Ashby et al.,~resubmitted). SEDS reaches a point-source sensitivity of $26$ AB 
mag ($3\sigma$) at both $3.6$ and $4.5\mu$m . 

Additional IRAC observations of parts of each SEDS field are now underway as part of the 
SEDS-CANDELS program (PI: G. Fazio). The S-CANDELS observations of the UDS (PID 80218), 
when completed, will cover $\sim$150~arcmin$^2$ within the CANDELS area and reach a maximum 
depth of about $50$ hours (four times the existing depth).

 \subsection{Spectroscopy}

A number of spectroscopic observations were conducted in the UDS field. They often targeted specific 
types of objects: passively evolving galaxies \citep{Yamada2005}, radio sources (Simpson et al. 2006, 2012; 
Vardoulaki et al. 2008; Akiyama et al. in prep.; Pearce et al. in prep.)\nocite{Simpson2006, Simpson2012, 
Vardoulaki2008}, QSOs \citep{Smail2008}, Ly$\alpha$ emitters \citep{Ouchi2008}, galaxy cluster members
 \citep{Geach2007, VanBreukelen2007, Papovich2010, Tanaka2010, Finoguenov2010}. 
 
Extensive spectroscopic campaigns have recently increased the number of known redshifts in the UDS 
field. A spectroscopic campaign using Magellan/IMACS multi-object spectrograph (PIs: M. Cooper \& 
B. Weiner) provides reliable spectroscopic redshifts for $272$ sources within the CANDELS UDS field of view 
(Cooper et al. in prep.\footnote[9]{http://mur.ps.uci.edu/\~cooper/IMACS/home.html}). 
The UDSz on-going ESO Large Programme (PI: O. Almaini) is being carried out with the VLT/VIMOS and 
VLT/FORS2 spectrographs and is obtaining spectra for $\sim3500$ sources in the full UDS field. Most of 
these spectroscopic redshifts are still proprietary, however, and will be added to the CANDELS catalog 
as soon as they become publicly released. At the time of publication, only $210$ sources in the catalog 
have a non-proprietary spectroscopic redshift.


\section{The CANDELS UDS $F160W$-selected catalogue}

\subsection{$F160W$ detection band}

\begin{figure}
\begin{center}
\includegraphics[width=7.5cm]{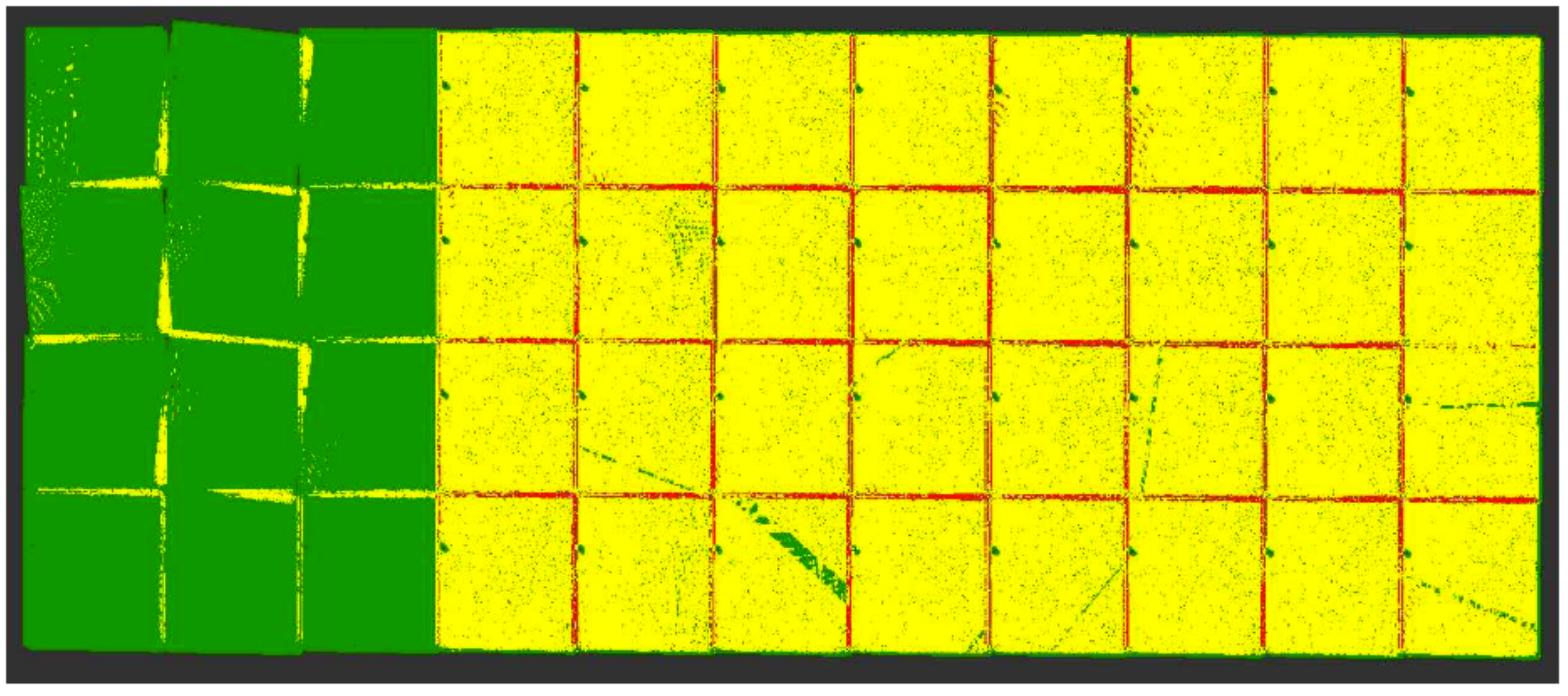}
\includegraphics[width=7.5cm]{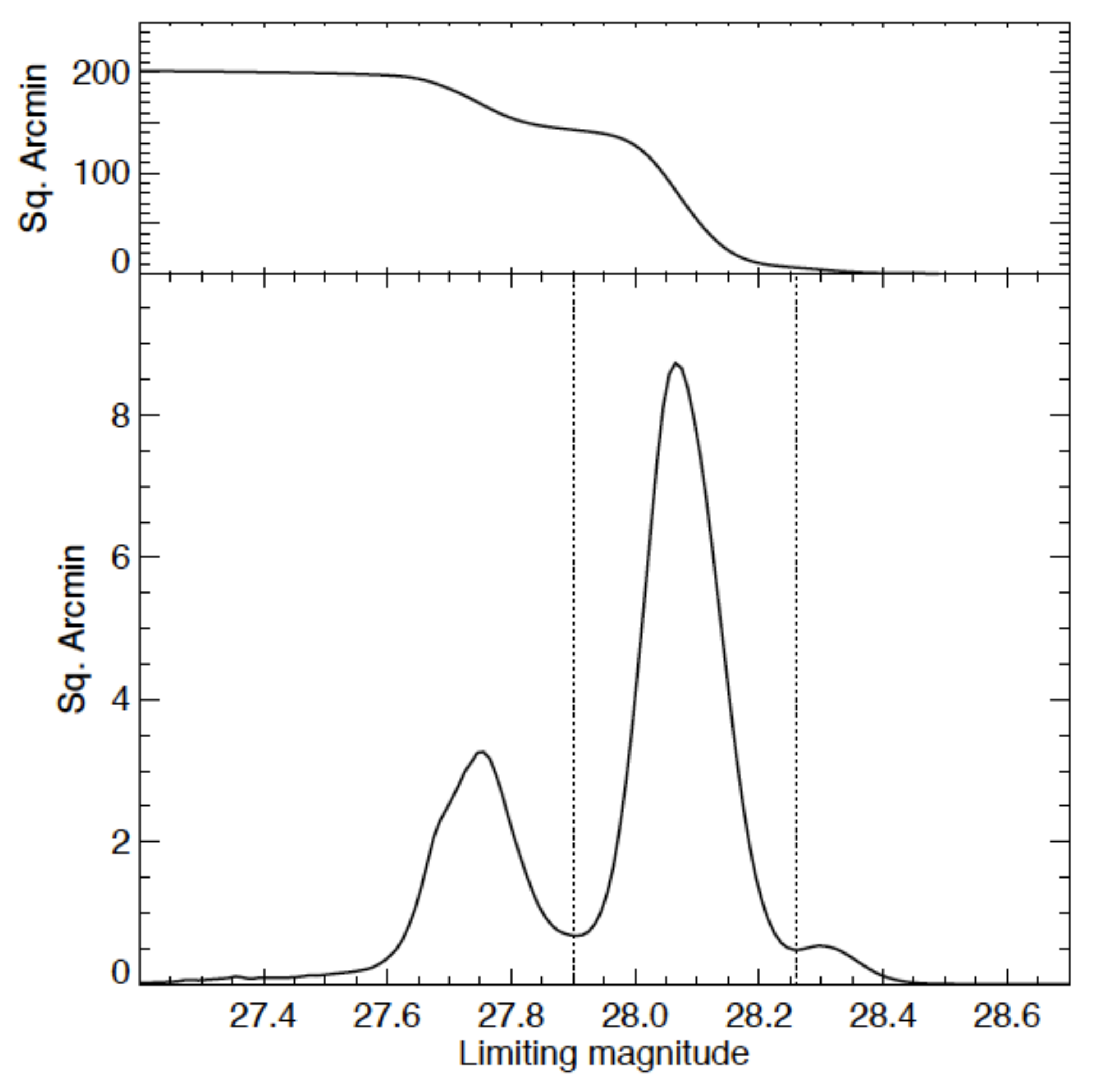}
\end{center}
\caption{Limiting magnitudes of the $F160W$ image within an area of $1$~arcsec$^2$ square
($1\sigma$). {\it Top:} Distribution over the CANDELS UDS field. Regions with limiting 
magnitudes mag $< 27.90$, $27.90 <$ mag $< 28.26$ and mag $> 28.26$ are shown in green, 
yellow and red respectively. {\it Bottom:} Distribution of area with a given limiting magnitude (per bin 
of $0.01$). The top inset shows the cumulative distribution of area with sensitivity greater than a 
given limiting magnitude.}
\label{limmag}
\end{figure}

The field of view covered by the $F160W$ data is about $201.7$ arcmin$^2$. Exposures are shorter at 
the eastern end of the mosaic (to accommodate $F350LP$ exposures during the same orbit, see 
Koekemoer et al.~2011, Figure~17). We estimate the limiting magnitudes as a function of position 
(from the rms map). They 
are computed for each pixel rescaled to an area of $1$~arcsec$^2$ at a $1\sigma$ level. 
Figure~\ref{limmag} shows the distribution of the limiting magnitudes across the $F160W$ image and
its associated histogram. The peaks in the distribution are found at $27.75$, $28.06$ and $28.30$ and
we therefore clearly identify three main areas at different depths: mag $< 27.90$ (the $\sim1/3$ shallower 
Eastern tiles), $27.90 <$ mag $< 28.26$ (the $2/3$ deeper Western tiles and tile overlaps in the shallow 
region) and mag $>28.26$ (tile overlaps in the deeper region) respectively corresponding to regions 
of about $58.7$, $136.9$ and $5.9$~arcmin$^2$. The present catalog also contains for 
each source an estimation of the limiting magnitude at the position where the source is detected 
(see \S3.3 for details).

\subsection{Source extraction}

\subsubsection{Modifications of the SExtractor software}

We used a slightly modified version of the SExtractor software version 2.8.6 \citep{Bertin1996} for the source 
extraction in the $F160W$ image.

{\it (i)} Several tests by the GOODS team\footnote[10]{\tt http://www.stsci.edu/science/goods/catalogs/r1.0z\_readme} 
showed that the `inner annulus'  adopted by SExtractor and used for the determination of the local 
sky background barely reaches the edges of sources (especially for faint sources), resulting in the wings 
of galaxies being included in the sky measurement and the total SExtractor source flux underestimated. 
We therefore adopt the same modification of the SExtractor code that ensures that the `inner annulus' 
is at least $1\arcsec$ radius wide \citep{Giavalisco2004, Grazian2006B}.

{\it (ii)} After the detection is performed, SExtractor runs a cleaning process that attempts to merge sources
that were falsely split. The original SExtractor cleaning function had a tendency to merge `non detection' 
sources to a real source close-by. We adopt a modified cleaning routine (i.e.~modified {\tt clean.c}) that ensures
that these sources are discarded.

{\it (iii)} We also modified SExtractor in various places to ensure that, in dual mode, it uses the gain of the 
measurement image, and not the detection image, when calculating isophotal-corrected magnitudes.

\subsubsection{SExtractor cold and hot detection modes}

\begin{figure*}
\begin{center}
\includegraphics[width=9cm]{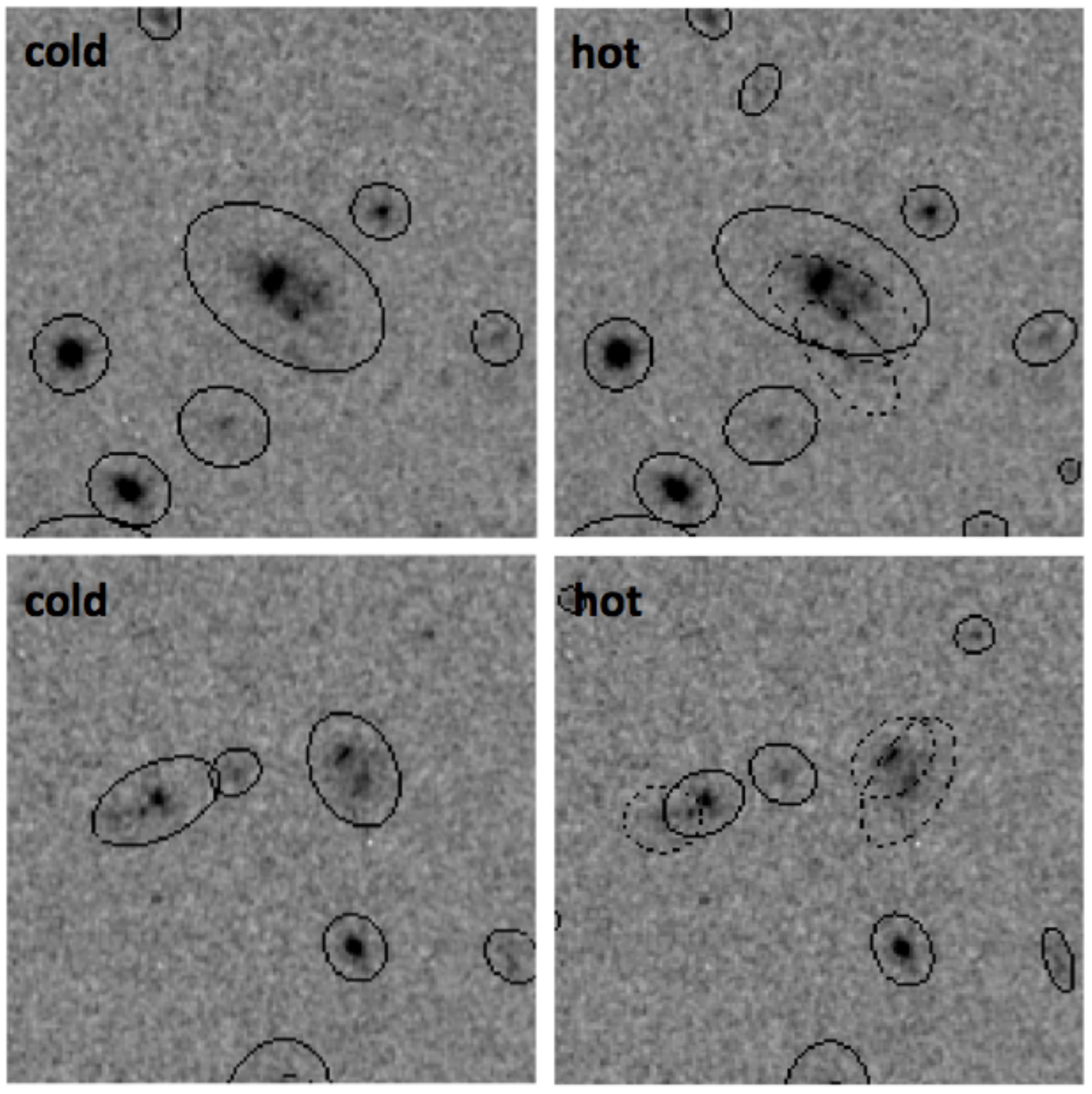}
\end{center}
\caption{SExtractor cold and hot modes. Left and right panels respectively show the 
differences between the cold and hot source extraction modes for two distinct regions 
(upper and lower panels; $10\arcsec \times 10\arcsec$) of the WFC3 $F160W$ data.
SExtractor Kron ellipses are overlaid. The first region is centered on a bright ($H\sim23$)
extended galaxy. The second region contains several clumpy galaxies ($H\sim25$).
In both cases, the faint sources that are only detected in the hot mode have $H\sim27$.} 
\label{coldhot}
\end{figure*}

Traditional 
wide field surveys were usually not very deep. The source extraction was therefore focussed on deblending 
bright, extended sources, avoiding for example, the separation of galaxy sub-structures into multiple objects. 
For deeper surveys, source extraction was usually tuned to push the detection in order to pick up faint and 
small galaxies. In contrast, recent surveys, and in particular CANDELS, reach unprecedented depth over 
wide areas. The scientific goals of such surveys extend from the study of nearby sources to the discovery of 
the farthest and faintest galaxies in the early Universe. It has been unfeasible to find a single setup 
for the extraction of sources. We therefore need to adapt the detection methodology. Recent works have 
adopted a rather simple strategy to deal with this issue, using a two-step approach 
\citep[see e.g.,~][and references therein]{Gray2009}. First, SExtractor is run in a so-called `cold' mode that 
extracts and deblends efficiently the brightest sources. It is then run in a so-called `hot' mode optimized for 
the detection of faint objects. The parameters were further adjusted to obtain a consistent photometry between 
the cold and hot mode. The adopted SExtractor cold and hot mode parameters are provided in Appendix A. 

Figure~\ref{coldhot} illustrates the differences between the cold and hot source extraction modes on the $F160W$ 
image in the vicinity of several extended, clumpy galaxies. The cold mode detects the clumpy galaxy as one single 
object, merging all the clumps together in one single source, while the hot mode tends to separate clumps into
individual objects. Nevertheless, the hot mode detects the faintest objects of the image that were missed by the cold 
mode. By making use of these two complementary detection modes, we aim at producing the most reliable and 
complete source catalog possible. SExtractor was run twice to create the cold and hot mode catalogs and associated segmentation maps. The 
cold and hot mode catalogs contain $27167$ and $37715$ sources respectively. 

\subsubsection{Cold + Hot combination routine}

\begin{figure*}
\begin{center}
\includegraphics[width=9cm, ]{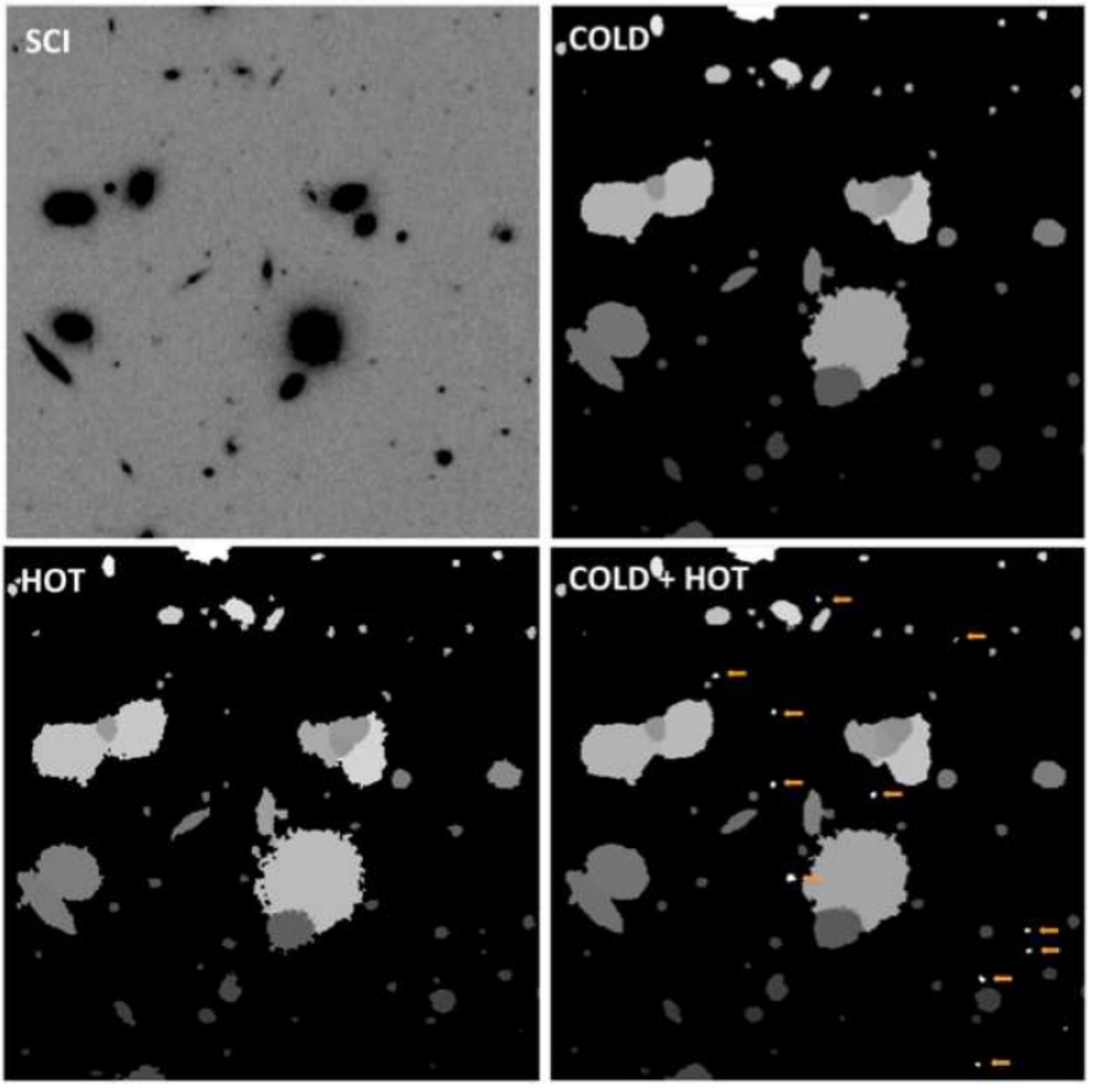}
\end{center}
\caption{Cold + hot combination routine. Panels respectively show a region of $40\arcsec \times 40\arcsec$ 
of the $F160W$ image (`SCI'; top left), the SExtractor segmentation map for the cold mode (`COLD'; top right) and 
the hot mode (`HOT'; bottom left) and the cold + hot combined segmentation map (`COLD + HOT'; 
bottom right). The shades of grey correspond to the source identification number in the SExtractor catalogs. 
The orange arrows, in the bottom right panel, indicate sources that were only detected in the hot mode and 
added to the cold mode catalog.}
\label{coldhotcomb}
\end{figure*}

The cold and hot catalogs were merged. The adopted cold + hot combination routine was adapted 
from GALAPAGOS, a software designed for source extraction, light-profile modeling and catalog 
compilation of large astronomical surveys (see Barden et al.~2012 and the GALAPAGOS manual for 
details\footnote[11]{\tt http://astro-staff.uibk.ac.at/$\sim$m.barden/galapagos/}).\nocite{Barden2012}

The combined cold + hot catalog includes all the sources detected in the cold mode + the sources 
detected in the hot mode at positions where no cold source was detected. In short, for each source 
in the hot mode, the routine checks whether it falls within the Kron ellipse of a cold mode 
detected source. If it does, the source is ignored. The final merged catalog therefore contains the full 
cold mode catalog followed by the `hot mode only' sources (with new updated identification indices). 

For example, in Figure~\ref{coldhot}, the routine will consider the different clumps of the extended 
galaxies as part of the same object --- because they are all detected within the same Kron ellipse 
in the cold mode --- and therefore appear as one unique object (one line) in the merged catalog (whose 
SExtractor characteristics are reproduced from the cold mode catalog). Sources that are only detected in 
the hot mode and isolated (i.e.,~not within a Kron ellipse of a cold mode source) are simply added to the 
catalog with the SExtractor measurements of the hot mode catalog.

Figure~\ref{coldhotcomb} 
shows a region of the $F160W$ image with the SExtractor segmentation maps of the cold, 
hot, and cold + hot modes. In the final combined segmentation map, the isophotal areas for sources detected 
in the cold mode are directly reproduced from the cold mode segmentation map. The isophotal areas for 
sources only detected in the hot mode are then added with their new identification number from the cold + hot 
merged catalog.

\begin{figure}
\begin{center}
\includegraphics[width=8.5cm, bb=0 50 550 450]{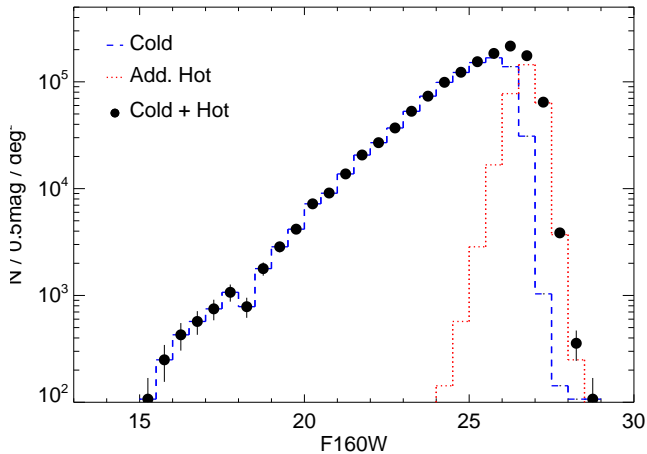}
\end{center}
\caption{Source number counts in the F160W band (black dots). Blue dashed and red dotted histograms 
respectively show the contribution of the cold and hot mode. We assume Poissonian error bars.}
\label{counts}
\end{figure}

The final combined $F160W$-selected catalog contains $35932$ sources. Figure~\ref{counts} 
shows the source number counts (per bin of $0.5$~mag; also listed in Table~\ref{counts2}) with the 
contributions of the cold and hot mode overlaid. The flux densities and uncertainties in the 
$F160W$-band reported in the multiwavelength catalog correspond to the SExtractor outputs 
FLUX\_BEST and FLUXERR\_BEST, converted into $\mu$Jy using the \citet{Koekemoer2011} zeropoints. 

The completeness limit of the UDS $F160W$ catalog was estimated by comparing with deeper data in the 
GOODS-South field. The central part of this field was observed in $10$ epochs (Grogin et al.~2011) i.e.,~five 
times the depth of UDS. 
We ran SExtractor (cold + hot) on the 10-epoch GOODS-S combined $F160W$ mosaic and derived number counts 
similarly to the UDS. By comparing number counts in GOODS-S Deep with
UDS, we estimate that the $90$\% ($50$\%) limit of completeness of the UDS $F160W$ data is $26.70$ ($27.05$).

\begin{table}
\caption{Number counts in the CANDELS UDS $F160W$ image}
\label{counts2}
\centering
\begin{tabular}{c c}
\hline
Mag		&	N mag$^{-1}$ deg$^{-2}$$^{\mathrm{a}}$\\
\hline 
$15.75$	&	$250\pm94$   \\
$16.25$	&	$428\pm124$   \\
$16.75$	&	$571\pm143$   \\
$17.25$	&	$749\pm164$   \\
$17.75$	&	$1071\pm195$   \\
$18.25$	&	$785\pm167$   \\
$18.75$	&	$1784\pm252$   \\
$19.25$	&	$2855\pm319$   \\
$19.75$	&	$4176\pm386$   \\
$20.25$	&	$7209\pm507$   \\
$20.75$	&	$9101\pm570$		\\
$21.25$	&	$13740\pm700$	\\
$21.75$	&	$20664\pm859$   \\
$22.25$	&	$26946\pm981$   \\
$22.75$	&	$36974\pm1149$   \\
$23.25$	&	$53070\pm1376$   \\
$23.75$	&	$73449\pm1619$   \\
$24.25$	&	$99003\pm1880$   \\
$24.75$	&	$122950\pm2095$   \\
$25.25$	&	$154357\pm2347$   \\
$25.75$	&	$184443\pm2566$   \\
$26.25$	&	$216207\pm2778$   \\
\hline     
\end{tabular}
\begin{list}{}{}
\item[$^{\mathrm{a}}$] Uncertainties on the number counts are Poissonian.

\end{list}
\end{table}

\subsection{Additional $F160W$-related values} 

$\bullet$ $F160W$ Limiting magnitude: For each source in the catalog, we derive a limiting magnitude 
indicating the depth of the $F160W$ image in the region where the source falls. This limiting magnitude 
is derived from the square root of the average of the rms squared in the SExtractor segmentation map of 
each source scaled to an area of $1$~arcsec$^2$. In practice, the limiting sensitivity in $HST$ images 
depends on the source size, and faint galaxies can be significantly smaller than $1$ arcsec$^2$ and 
thus be detected fainter than this fiducial magnitude limit. The knowledge of the depth fluctuation of the 
detection image is fundamental for any future volume-sensitive statistics (e.g.,~luminosity functions etc.). 

$\bullet$ Flag: Not all objects detected by SExtractor are real. For example, SExtractor typically detects spurious 
sources in the spikes of very bright stars. Some sources also fall at the edges of the detection image where 
the photometry is not optimal. A flag column is therefore included in the catalog which indicates star spikes 
and bright halos as well as large artifacts and noisy edges (see Appendix B for details).

\section{Photometry of the HST data}

The photometry of the other {\it HST} bands (i.e.,~WFC3 $F125W$, ACS $F606W$ 
and $F814W$) was derived using SExtractor in dual-mode with a source detection on the $F160W$ image. 
To take into account the PSF differences between the F160W detection image and 
the $F125W$, $F606W$ and $F814W$ images, these measurement images are 
PSF-matched to the $F160W$ data.  Empirical PSFs were derived from stacking the 
images of several isolated and unsaturated stars in the field.  In order to provide a 
more accurate description of the central region, we replaced the inner-most  
pixels (within a radius of 3 pixels from the center) with a simulated PSF generated 
with the {\tt TinyTim} package (Krist 1995)\nocite{Krist1995}.  The {\tt TinyTim} PSF was dithered and drizzled 
in the same manner as the observations, and normalized such that the total flux of 
the newly constructed hybrid PSF model is the same as that of the stacked star.  
We found this hybrid PSF accurately reproduced the growth curves of stars out to $3$\arcsec.  
Further details on the PSF models can be found in \citet{VanDerWel2012}. Using 
{\tt iraf/psfmatch}\footnote[12]{IRAF is distributed by the National Optical Astronomy Observatory, 
which is operated by the Association of Universities for Research in Astronomy (AURA) under 
cooperative agreement with the National Science Foundation.}, we generated matching kernels 
from these hybrid PSFs, replacing the 
high-frequency and low signal-to-noise components of the PSF matching function 
with a model computed from the low frequency and high signal-to-noise components.
Similarly to the $F160W$ catalog, SExtractor was then run twice for each band to create the 
cold and hot-mode catalogs that were then merged.

Part of the $F160W$ field of view (i.e.,~the eastern $\sim1/4$ of the CANDELS UDS field; 
see Figure~\ref{area}) was not covered by ACS. For sources detected in $F160W$ but 
outside the ACS field of view, we set the flux density and uncertainties for $F606W$ 
and $F814W$ to $-99$.
The region covered by both WFC3 and ACS is about $154$ arcmin$^2$. 

Although commonly adopted Kron magnitudes (SExtractor MAG\_AUTO) are mainly consistent with
isophotal magnitudes for bright or faint isolated sources, they are usually estimated over areas
larger than isophotal ones therefore resulting in a lower signal-to-noise. They can also be contaminated 
by neighboring sources. In high-resolution data such as {\it HST}, the isophotal area 
tends to follow more precisely the apparent size of the sources. Using isophotal magnitudes also 
guarantees that the flux is derived in the same area as the isophotal area defined by SExtractor for 
the $F160W$ image. For these reasons, isophotal magnitudes were adopted in past multiwavelength 
catalogs such as the GOODS-MUlticolor Southern Infrared Catalog (GOODS-MUSIC; Grazian et 
al.~2006)\nocite{Grazian2006B}, a multiwavelength catalog (visible to mid-infrared) that covers 
$143.2$ arcmin$^2$ in the GOODS-South field. Following these past works, we adopt isophotal 
colors for all sources assuming that the ratio of FLUX\_BEST $/$ FLUX\_ISO fluxes represents 
the aperture correction to total flux in F160W, and apply this uniformly to all HST bands. SExtractor 
isophotal flux densities (FLUX\_ISO) and flux uncertainties (FLUXERR\_ISO) for $F125W$, $F606W$ 
and $F814W$ were first converted into $\mu$Jy using \citet{Koekemoer2011} zeropoints and then 
into total flux densities following:\\

{\noindent}$F(\lambda) = F_{Iso}(\lambda) \times F_{Best}(160) / F_{Iso}(160)$\\
$F_{unc}(\lambda) = F_{unc; Iso}(\lambda) \times F_{Best}(160) / F_{Iso}(160)$

\section{Photometry for the ground-based and {\it Spitzer} images}

\begin{figure*}
\begin{center}
\includegraphics[width=12cm]{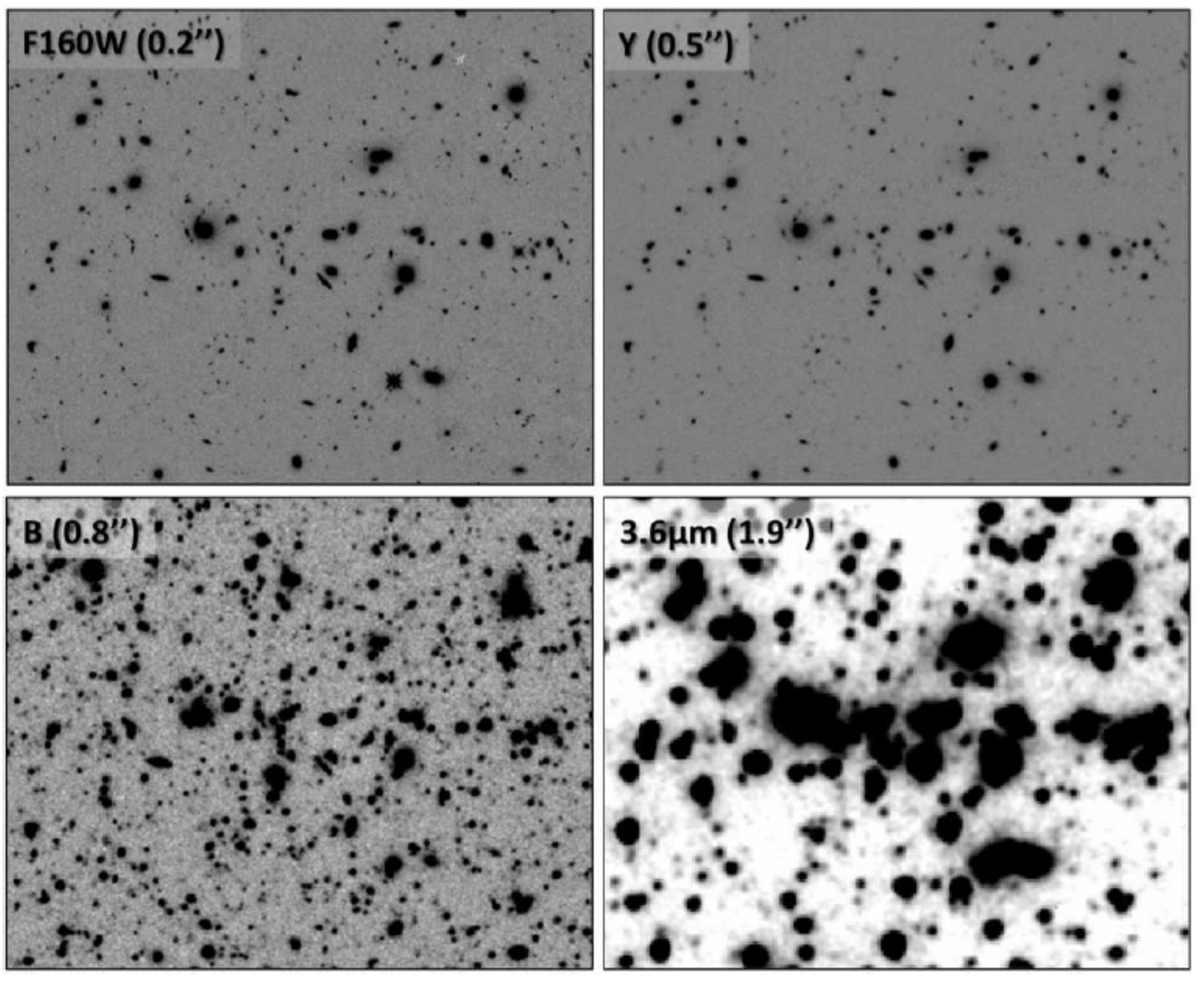}
\end{center}
\caption{$2\arcmin \times 1\farcm5$ region of the CANDELS UDS field seen in four of the bands 
available (from high to low-resolution data): the CANDELS HST $F160W$ (top left), VLT/HAWK-I $Y$ (top right), the
Subaru $B$-band (bottom left) and {\it Spitzer}/IRAC SEDS $3.6\mu$m (bottom right). The 
FWHM of each band is also indicated (in parentheses). This particular crowded region of the UDS field 
hosts a galaxy cluster at $z = 0.65$ \citep{Geach2007}.}
\label{psf}
\end{figure*}

It has always been challenging to derive reliable photometry for multiwavelength imaging surveys,
which combine data with a large variety of survey depth and resolution. The CANDELS UDS 
field is a typical example, covered by both high-resolution HST data and lower resolution 
ground-based and {\it Spitzer}/IRAC data. Figure~\ref{psf} shows the same region of the CANDELS
UDS field in HST $F160W$, VLT/HAWK-I $Y$, the Subaru $B$-band and {\it Spitzer}/IRAC SEDS 
$3.6\mu$m. The field of galaxies can appear to be quite different through different bands, due to their 
varying colors, the different depths of the data sets, and particularly the different PSFs. Close neighbors 
that appear isolated in high resolution data can appear strongly blended at lower 
resolution. It is therefore crucial to be able to deblend these sources from their projected neighbors.  

Several efforts were conducted to overcome these issues with the development of optimized photometry 
software \citep[e.g.,~{\tt ConvPhot};][]{DeSantis2007}. In the present work, we make use of the template-fitting 
photometry software TFIT, for consistency with the other CANDELS multiwavelength catalogs (e.g.,~Guo et 
al.~submitted for GOODS-S), to derive the photometry for all the non-{\it HST} data. TFIT has the advantage 
of using shifted kernels to 
account for any remaining small image distortion in the low-resolution images. It also works on original 
pixel scale of the low-resolution image. \citet{Laidler2007} and \citet{Papovich2001} 
provide a complete description of the TFIT software and \citet{Lee2012} present a set of simulations that validate 
this template-fitting technique and quantify its uncertainties. We only summarize the main steps in the following. 
In brief, TFIT uses {\it a-priori} information on the position and surface brightness profile of sources measured on a 
high resolution image (in the present work, the $F160W$-band) as priors to derive their corresponding photometry in 
lower-resolution images. As mentioned earlier, one of the main reasons for using TFIT is to derive reliable photometry 
for sources that overlap in the low-resolution image. TFIT uses information from their high-resolution (often well separated) 
counterparts and fits them simultaneously. First, TFIT builds a low resolution (normalized) template model of each 
source by smoothing the high resolution image of the object to the PSF of the low-resolution image using a convolution 
kernel. The best fit fluxes to the low resolution image are then derived from these templates using chi-square minimization. 

\subsection{Preparing the low-resolution images}

TFIT requires a careful preparation of the low-resolution images.

$\bullet$ Astrometry: The low-resolution images should also have a compliant astrometry with the high-resolution 
images. The pixel scale of the low-resolution image should be equal to (or an integer multiple of) the pixel scale of the 
high-resolution image. All ground-based images were resampled to the $F160W$ pixel scale of $0.06$\arcsec 
and aligned to the CANDELS {\it HST} data astrometry using {\tt swarp} \citep{Bertin2010}. 
We keep the original pixel scale ($0.6$\arcsec) of the {\it Spitzer} data (SEDS, SpUDS). The astrometry 
of SEDS (Ashby et al.~resubmitted) is compliant to the astrometry of the HST images so no realignment 
was needed.  This was not the case for the SpUDS data that were therefore reprojected to the HST 
astrometry also using {\tt swarp}. 

$\bullet$ Background subtraction: The low-resolution images first need to be background-subtracted. 
For each image, a first
rough background approximation was determined by smoothing the image on large scales using a 
large-annulus ring-median filter and then subtracted from the image. The image was then smoothed to 
the corresponding image PSF scale and sources were masked. To account for the wings of the sources,
the source masks were enlarged by convolving with a Gaussian kernel. Several iterations were performed, 
starting from the brightest sources with high signal-to-noise (S/N) ratio down to sources with a S/N$\sim5$ 
above the mean background. The non-masked pixels were used to create a `noise' map, which was interpolated
to determine the background in the masked pixels. 

We tested the background subtraction routine by sampling the noise in the background-subtracted low-resolution 
images. We randomly added $1000$ artificial sources in the $F160W$ image (privileging regions without 
sources) and derived the photometry of their non-existent counterparts in the low-resolution images. We 
expect the S/N (flux density over uncertainty) to be a Gaussian, centered to zero with $\sigma=1$. The 
background subtraction technique works well for ground-based data. 

For IRAC data, however, both the mean S/N and the TFIT residual image (see \S5.4) are found to be slightly
negative ($\sim10^{-4}$~MJy/sr) i.e.,~the routine is over-subtracting the background. Although this
quantity is negligible for bright objects, it becomes dominant for faint sources. We therefore implemented an
additional background correction for the IRAC data. We ran TFIT on the background-subtracted images,
subtracted the TFIT output collage --- an image of the sources as modeled by TFIT --- and repeated the 
background subtraction procedure. We then repeated the test with artificial sources and found a 
Gaussian-like distribution of the S/N and mean residuals $< 10^{-6}$~MJy/sr). TFIT was run a fourth time
on the final image.

\subsection{Preparing kernels and PSF}

One of the key processes within TFIT is the smoothing of the high resolution detection image to the PSF of the 
lower resolution measurement image. Such a smoothing process is performed by applying a convolution 
kernel. The derivation of the convolution kernel is not done by TFIT and must be provided beforehand by the users. 

Several techniques have been developed to derive an accurate convolution kernel, 
in particular those by \citet{Alard1998} and \citet{Alard2000}.
In the case of the UDS visible and near-infrared ground-based data, we adopt a slightly simpler method based on 
analysis in the Fourier space, similar to the convolution kernel derivation technique presented in \citet{Aniano2011}. 
Such method can be implemented with standard astronomical tools for image analysis and has been successfully 
used for the GOODS-MUSIC \citep{Grazian2006B, Santini2009} and GOODS-ERS database \citep{Santini2012}.
In brief, we first derive the PSFs of the detection and measurement image, $PSF_1$ and 
$PSF_2$ respectively, e.g.,~by summing up stars in the field.
These two PSFs are then normalized. 
The convolution kernel $K$ is, by definition:

\begin{equation}
PSF_2(x,y)=K(x,y) \otimes PSF_1(x,y)
\end{equation}

The derivation of the exact shape of the convolution kernel is done in the Fourier space i.e.~if 
${\aleph} = FT(K)$, ${\wp}_1=FT(P_1)$ and ${\wp}_2=FT(P_2)$, the Fourier transform of the 
kernel is given by:
\begin{equation}
{\aleph}=\frac{{\wp}_2}{{\wp}_1}
\end{equation}
 
A low passband filter (LPBF)\footnote[13]{Filter functional form: $f(x,y) = 1 / (1 + \alpha \times (D/D_o)^{2n})$ 
where $D$ is the Euclidean distance of the (x,y) point from the central coordinates. We varied the values of $\alpha$, 
$D_o$ and $n$ to find the best residual.} was applied in the Fourier domain to remove 
the effects of noise and suppress the high frequency fluctuations. 
A residual was computed from the two PSFs in order to check for 
the validity of the derived filter.
Finally, ${\aleph}$ was transformed back to the pixel space and normalized to unity (to preserve 
the flux of the objects) following:

\begin{equation}
K(x,y)=FT^{-1}({\aleph}\cdot LPBF)
\end{equation}

If the high and low-resolution images have significantly different resolutions --- such as {HST} and 
{\it Spitzer} --- the PSF of the low-resolution images can be used directly as the convolution kernel.
For IRAC, we obtained a set of $5 \times$ spatially-oversampled PSFs (W. Hoffman; private
communication; see also IRAC Data Handbook, Appendix C).
There are $25$ such oversampled PSFs for each channel, representing a $5 \times 5$ sampling
across the detector to map PSF variation.  We averaged together these $25$ location-specific
PSFs to construct a single, location-averaged, PSF for each channel. The SEDS and SpUDS images are mosaics 
of a series of Astronomical Observation Requests (AOR; $107$ and $26$ respectively) that were 
executed with different PAs. The model PSF is rotated by the PAs for each AOR. The average of all 
rotated PSFs is then smoothed by a boxcar kernel, a mandatory step because the model PSF is sharper than 
the true IRAC PSF. Tests on the SEDS and SpUDS data showed that the optimal smoothing is obtained 
using a boxcar size of $23$~pixels ($0.06\arcsec$/pixel). Finally, the smoothed PSF are circularized 
by putting the pixels outside a diameter of $25\arcsec$ to zero and normalizing the total `flux' in the 
template to unity.

\subsection{Preparing TFIT input source catalog and segmentation map}

\begin{figure}
\begin{center}
\includegraphics[width=8.5cm]{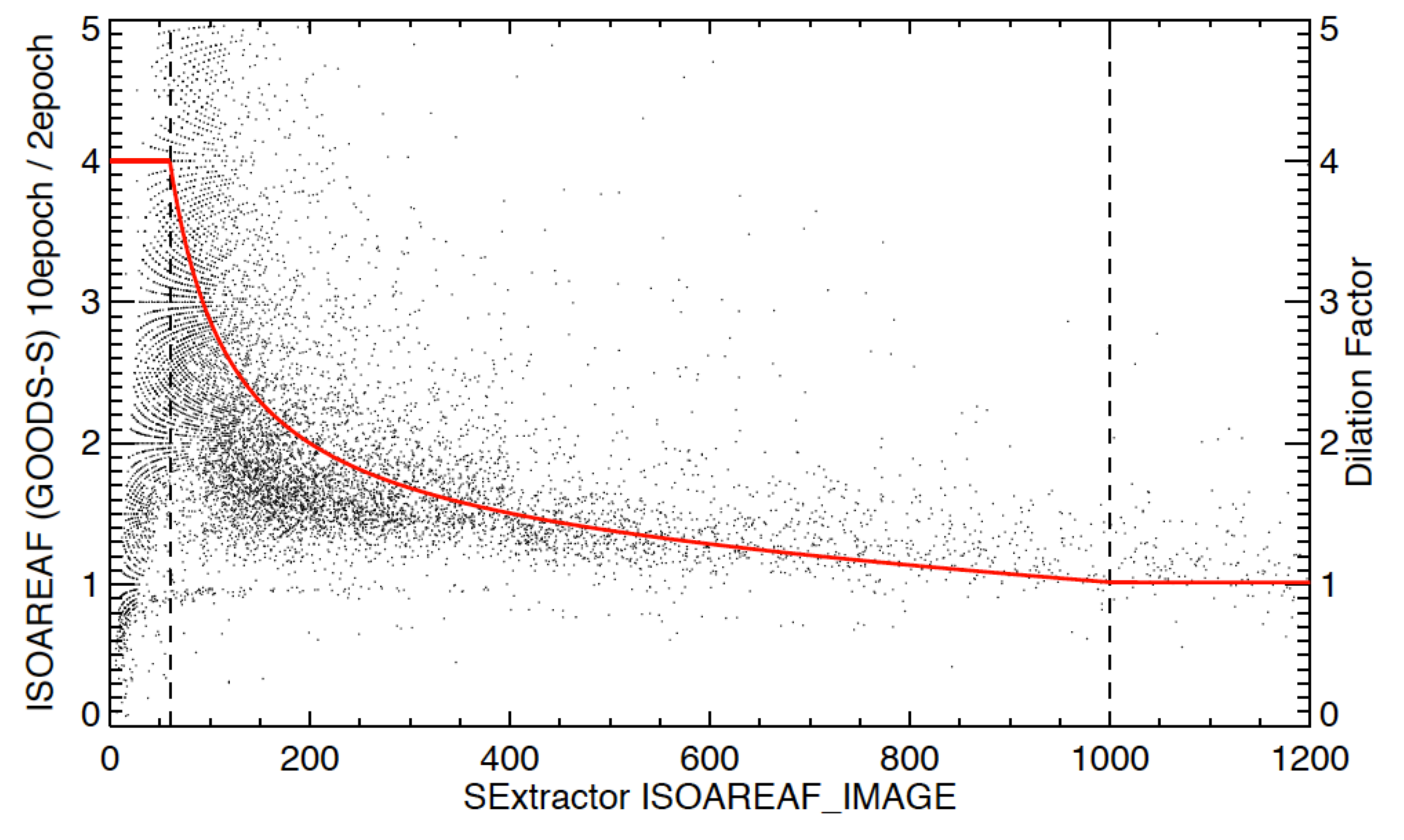}
\end{center}
\caption{Ratio between the $F160W$ SExtractor ISOAREAF\_IMAGE of sources in GOODS-South 
CANDELS Deep `10Epoch' (i.e.~full CANDELS Deep image) and `2Epoch' (i.e.,~after the two first 
CANDELS epochs) in function of the ISOAREAF\_IMAGE of the `2Epoch' (black dots). Also plotted 
are the dilation factor correction (red line) depending on a  given source isophotal area (vertical dashed lines; 
see \S4.3.2).}
\label{dilaf1}
\end{figure}

\begin{figure}
\begin{center}
\includegraphics[width=8.5cm]{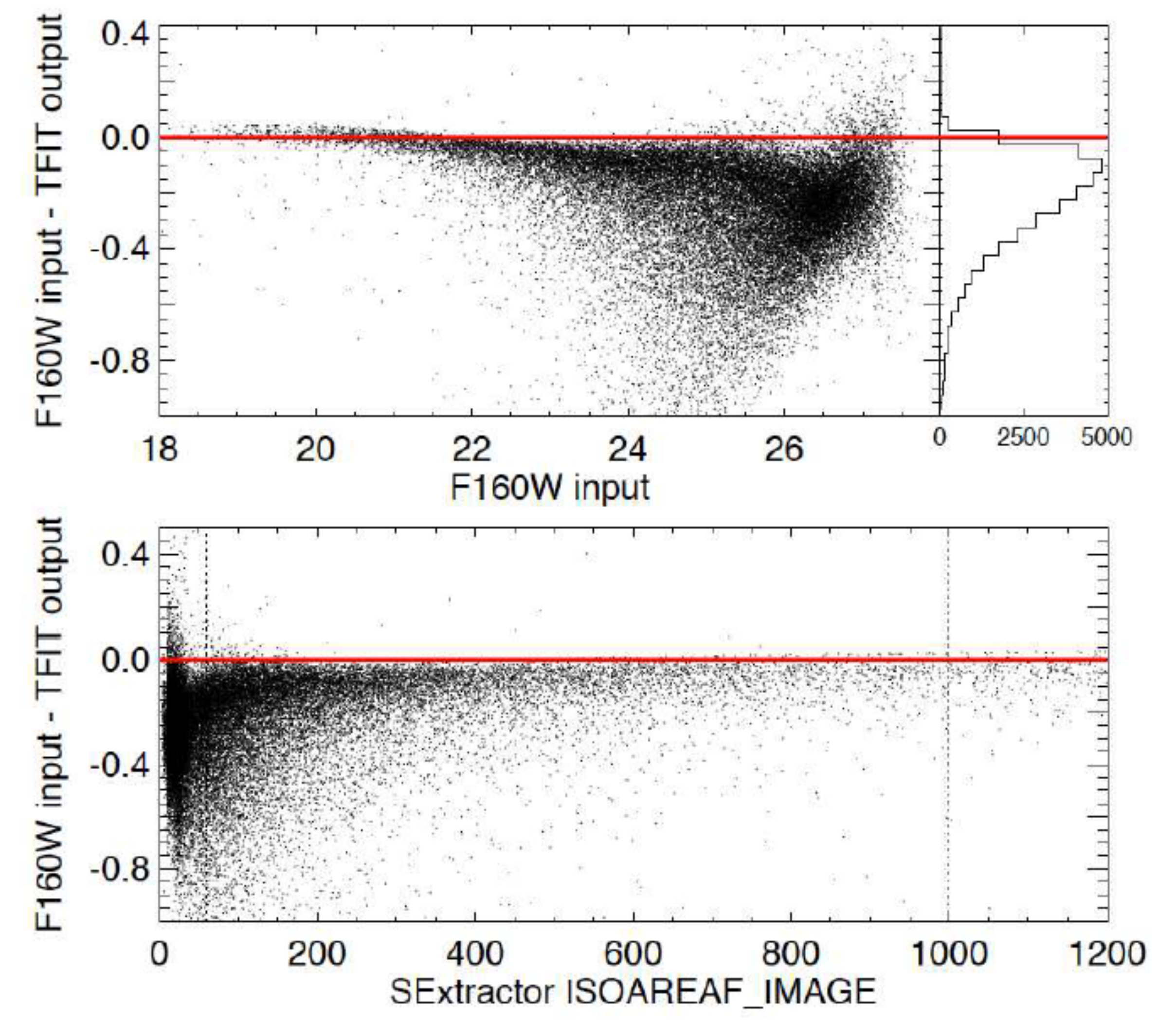}
\end{center}
\caption{
TFIT output photometry derived from the $F160W$ image smoothed to a resolution of $0.4\arcsec$ (i.e.,~to 
the resolution of the HAWK-I data): {\it Top:} versus the $F160W$ initial photometry (SExtractor MAG\_BEST). 
The histogram of the distribution is plotted in the right inset {\it Bottom:} versus the isophotal area (SExtractor 
ISOAREAF\_IMAGE). The vertical dotted lines mark the positions for ISOAREAF\_IMAGE $= 60$ and 
$1000$ that will be relevant for our dilation correction (see \S5.3.2). TFIT fluxes are gradually underestimated for 
fainter and/or smaller sources.}
\label{dilbf}
\end{figure}

\begin{figure}
\begin{center}
\includegraphics[width=8.5cm]{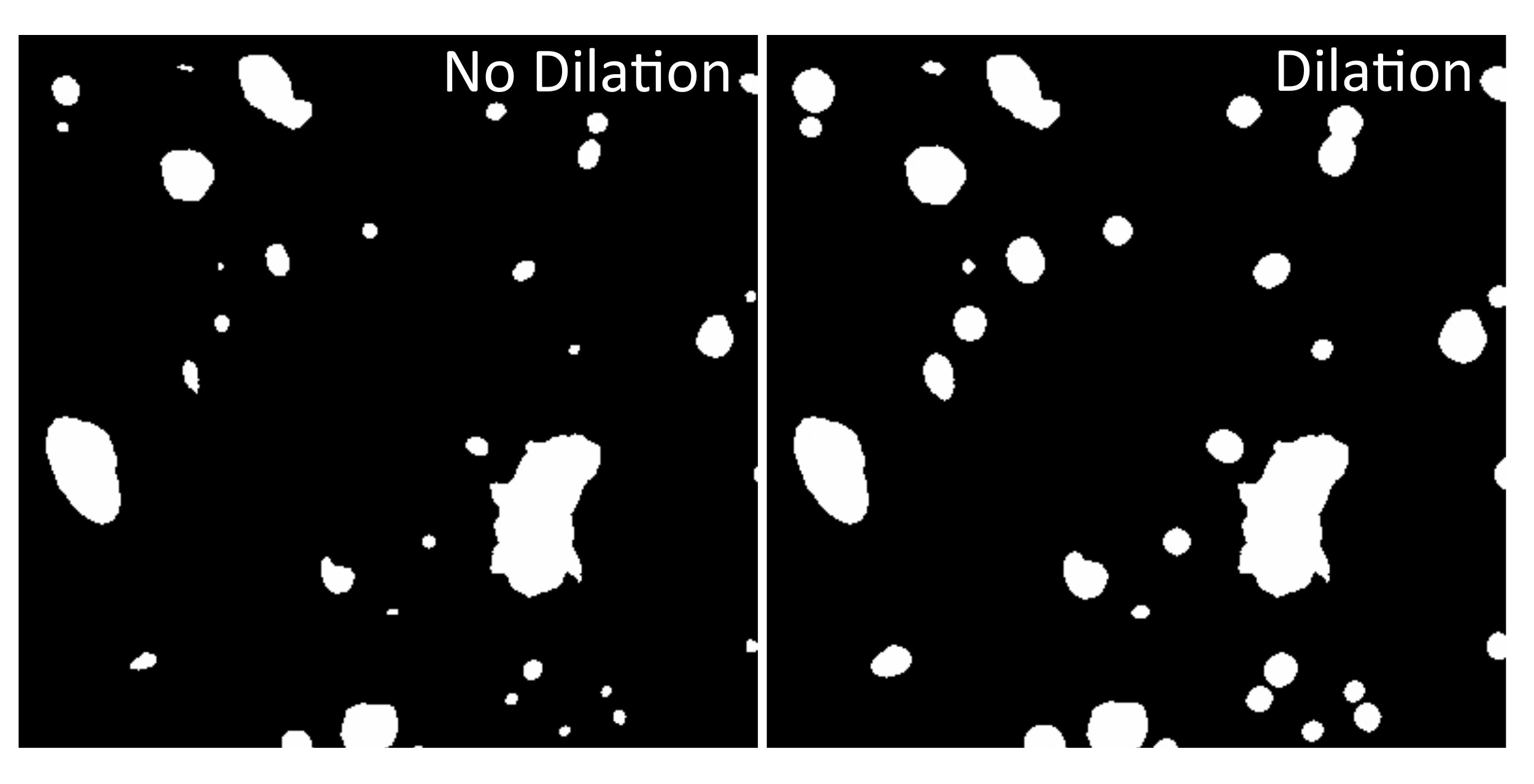}
\includegraphics[width=8.5cm]{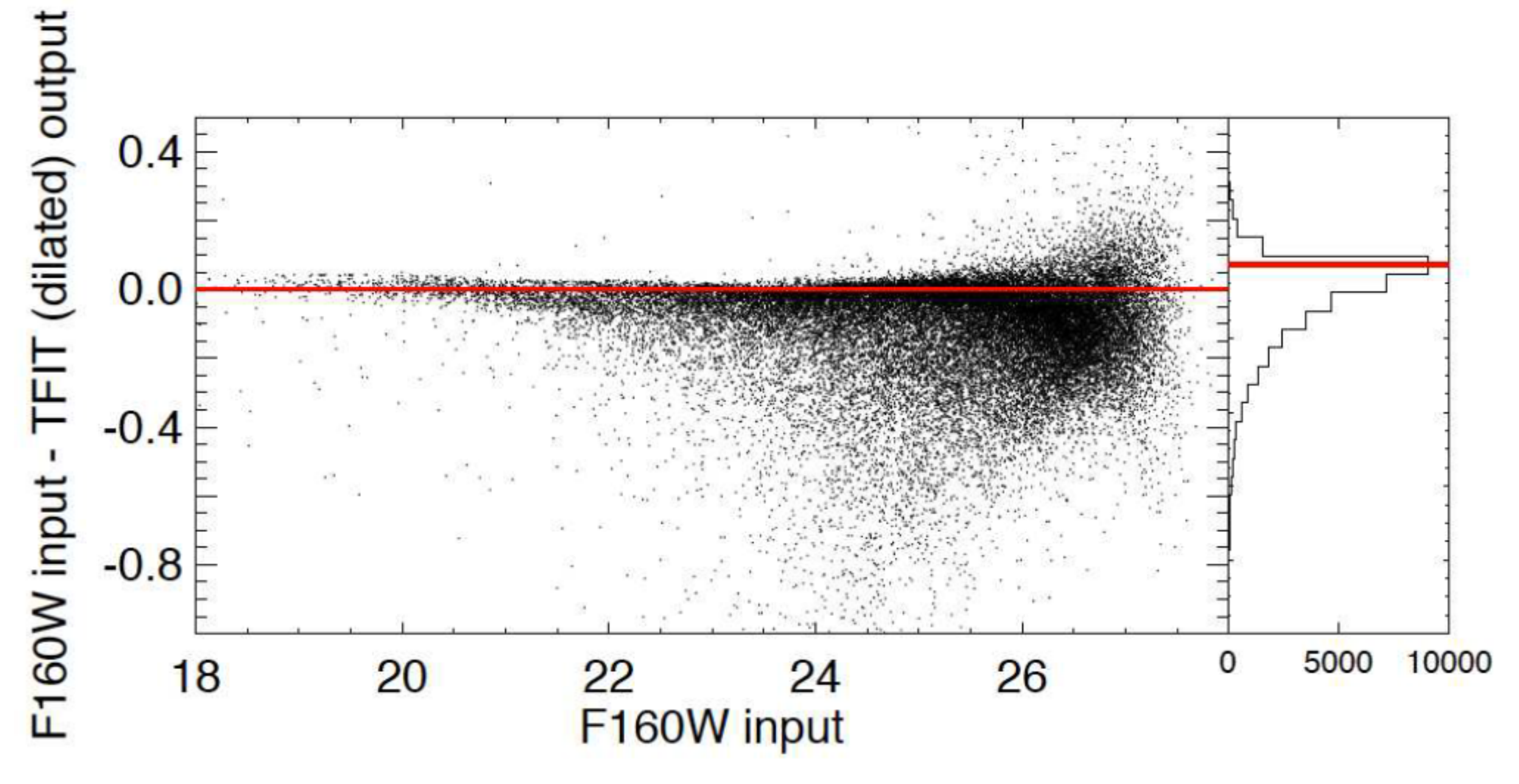}
\end{center}
\caption{Dilation correction. {\it Top:} Portion of the segmentation map before (left panel) and after dilation (right). 
Isophotal area larger than $1000$~pixels are not dilated whilst smaller areas are dilated according to the 
adopted correction (see text and Figure~\ref{dilaf1}) {\it Bottom:} Same as Figure~\ref{dilbf} but providing TFIT 
with SExtractor dilated input segmentation map. The TFIT output is now much more consistent with the $F160W$ 
input photometry thanks to the implemented dilation correction.}
\label{dilaf2}
\end{figure}

In order to derive photometry, TFIT uses as the source template only pixels within a segmented region. 
In past works using TFIT (e.g.~Laidler et al.~2007), this region was defined by the SExtractor segmentation 
map. This implies that only pixels within the HST isophotal area were used to construct templates for TFIT. 
However, past studies (e.g.~De Santis et al.~2007) have shown that SExtractor tends to underestimate 
the size of the source isophotal areas especially for faint and/or small sources and that therefore part of 
the source flux was missed. Unfortunately, it is challenging to estimate how much of the flux is missed exactly 
because {\it (i)} we do not know how much flux is actually outside the segmented region for each source, and 
{\it (ii)} a portion of this flux will be considered as background in the preparation of the low-resolution 
images for TFIT. The size of the isophotal area depends on the threshold adopted in the detection image. 
However, for faint sources, a significant fraction of their flux probably falls below the isophotal detection threshold.
This isophotal area is fed as an input parameter to TFIT which, as a consequence, tends to miss a significant 
fraction of the flux and underestimate the total magnitude.
We run a series of tests to quantify how SExtractor underestimates the isophotal area and implement 
a `dilation' correction to compensate for this underestimation. This issue mostly concerns sources that 
are either faint or with a small isophotal area, two populations that are overlapping as size strongly 
correlates with brightness. We will interchangeably refer in the following sections to faint/bright or 
small/large sources.

\subsubsection{SExtractor estimation of source isophotal area}

The following tests were done on the $F160W$ band since it is the catalog detection band and the reference 
high-resolution image for our TFIT run.

As mentioned in \S3.2.3, the central part of the GOODS-South field was observed in ten epochs, five times the 
UDS depth. We ran SExtractor (`cold + hot' then merged) on the full GOODS-S $F160W$ image ('10Epoch') and 
also on an image made only from the first two epochs (`2Epoch'). Figure~\ref{dilaf1} shows the ratio of 
ISOAREAF\_IMAGE parameter between the `10Epoch' and `2Epoch' images. SExtractor derives larger source 
isophotal areas (ISOAREAF\_IMAGE, the number of pixels above the adopted threshold) in the deeper image. As 
expected from previous studies, the isophotal area is smaller in the shallower image for faint sources. 
Although SExtractor nicely recovers the total ISOAREAF\_IMAGE for bright sources 
(ISOAREAF\_IMAGE $> 1000$~pixels), it gradually loses part of the source for fainter objects with an underestimation 
of at least a factor of $4$ for sources with ISOAREAF\_IMAGE $= 60$~pixels. 

As a test of how isophotal area affects photometry, we smoothed the $F160W$ image to the 0\farcs4 resolution of the 
VLT/HAWK-I data by convolving the image by the corresponding kernel. Using this 
same kernel, we ran TFIT, intending to recover the input photometry of the original (unsmoothed) $F160W$ 
image. Figure~\ref{dilbf} shows the difference between the output of TFIT and the input $F160W$ magnitude 
versus input magnitude (top) and SExtractor isophotal area (bottom; i.e.~the 
ISOAREAF\_IMAGE parameter). Figure~\ref{dilbf} shows that the TFIT output is in good agreement 
with the $F160W$ input values for bright sources ($mag_{F160W} < 20$ and ISOAREAF\_IMAGE $> 1000$~pixels) 
but worsens rapidly with faintness (and smaller isophotal area) with an average offset of $-0.24$ for $26 < 
mag_{F160W} < 27$ ($-0.26$ for sources with ISOAREAF\_IMAGE $< 60$~pixels). As expected, SExtractor 
tends to underestimate the isophotal area 
of sources and therefore TFIT fails to recover the expected input photometry for faint objects. 
Smoothing the $F160W$ data to match even lower-resolution data e.g.,~$0.8\arcsec$ (Subaru/Suprime-Cam) 
or {\it Spitzer}/IRAC gives trends similar to those of Figure~\ref{dilbf}, although the discrepancies between 
input and TFIT output are less dramatic with an average offset of $-0.1$ for $26 < F160W < 27$. 

\begin{figure*}
\begin{center}
\includegraphics[width=12cm]{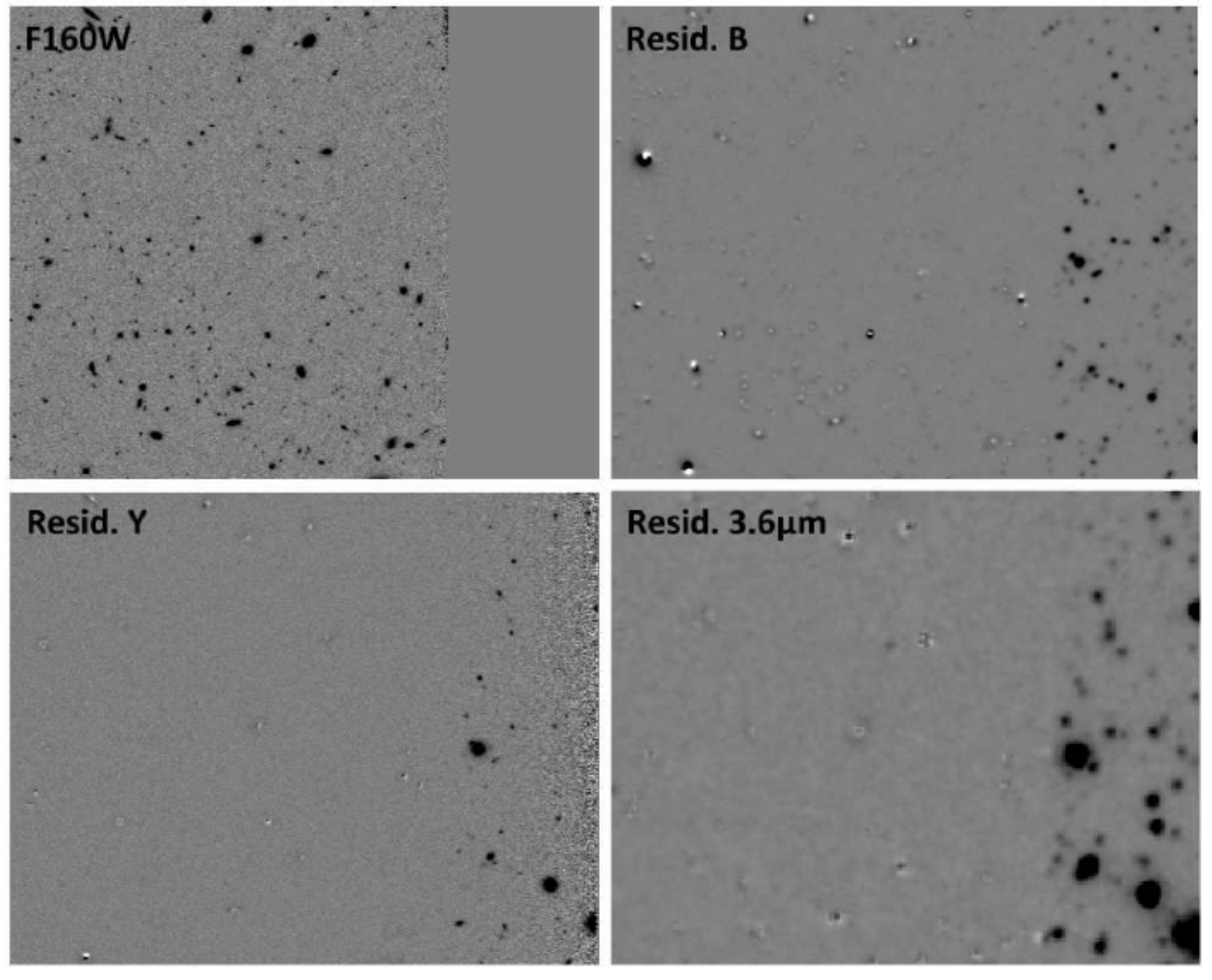}
\end{center}
\caption{$2.5\arcmin \times 1.5\arcmin$ sub-region of the TFIT residual output images. The first panel 
shows the $F160W$ detection image. We deliberately 
choose the edge of the $F160W$ field of view to visually assess the effect of the source removal by TFIT 
in the $B$-band (top right), the $Y$-band (bottom left) and SEDS $3.6\mu$m band (bottom right).}
\label{resid}
\end{figure*}

\subsubsection{Dilation correction}

In order to deal with this particular issue, we adopt a correction technique similar to the one implemented 
within the ConvPhot software. ConvPhot automatically dilates 
the area of every source on the segmentation map generated by SExtractor. 
Sources with an area above a minimum threshold ($m_{(AREA)}$) are dilated by a constant factor of $4$ 
(i.e.,~doubling the area). Sources smaller than $m_{(AREA)}$ are dilated to reach this minimum threshold. 
The dilation correction is done using a publicly available routine called {\tt dilate} \citep[see~][]{DeSantis2007}. 
{\tt dilate} expands the segmentation map by a fixed factor but prevents the merging between close sources.

We refine this dilation technique by modifying the {\tt dilate} routine to adapt it to the present dataset. As observed 
in Figure~\ref{dilbf} (as well as in Figure~\ref{dilaf1}), TFIT recovers well the photometry of sources 
with ISOAREAF\_IMAGE $> 1000$~pixels with a difference between the input and TFIT output magnitude 
of less than $< 0.01$. We therefore do not apply any dilation for these sources. The largest discrepancies are 
obtained for sources with an isophotal area smaller than $60$~pixels (with an underestimation of a factor of 
$4$ of the isophotal area in the case of GOODS-South `2Epoch'/`10Epoch' comparison). We therefore adopt
a constant dilation factor of $4$ for sources with ISOAREAF\_IMAGE $< 60$~pixels. As observed in Figure~\ref{dilaf1}, 
the loss of isophotal area by SExtractor can be parametrized by an hyperbola. We therefore opt for a smooth 
transition for sources with $60 <$ ISOAREAF\_IMAGE $< 1000$. To summarize, we adopt the following 
dilation factor corrections:\\
 
{\noindent}$\bullet$ If ISOAREAF\_IMAGE$>1000$: Dilated\_area$=$area\\
$\bullet$ If $60 <$ ISOAREAF\_IMAGE $< 1000$:\\
Dilated\_area $= -0.0004 \times$ area$^2 + 1.25 \times $area $+166$. \\
$\bullet$ If ISOAREAF\_IMAGE$< 60$: Dilated\_area$=4 \times$area\\

Figure~\ref{dilaf1} shows the adopted dilation correction. Figure~\ref{dilaf2} shows its effect on the 
segmentation map. It also shows the final discrepancies between the input $F160W$ photometry and 
the TFIT output after dilation. The adopted dilation correction permits us to recover, with more 
accuracy, the $F160W$ input photometry. 

This dilated segmentation map (and updated input source catalog) is adopted for all the TFIT 
runs on the low-resolution data (ground-based and {\it Spitzer}/IRAC data).

\subsection{TFIT}

TFIT incorporates a procedure that quantifies any small geometric 
distortions and shifts that could persist even after a careful alignment of the high- and low-resolution 
images and creates a series of shifted kernels based on these distortions 
\citep[as part of the `registration dance' stage;~see~][for details]{Laidler2007}. We ran TFIT a second 
time with these kernels in order to improve the alignment of the model with the data. The shifts are typically 
less than $0.1$~arcsec for ground based data and less than $0.3$~arcsec for the {\it Spitzer} images.
For background-subtraction purposes described in \S5.1, TFIT was run a third time on the IRAC data.

TFIT produces a residual image derived by subtracting the model image -- a collage of the
source template scaled by the TFIT flux measurement --- from the low-resolution image. The
residual image permits us to visualize any TFIT problem such as a non-optimal convolution kernel.
Objects that are detected in a low-resolution image but are not in the input catalog derived from
the high-resolution data also remain unchanged on this residual image. Residual images of the 
TFIT second pass are shown in Figure~\ref{resid} for the $B$, $Y$ and SEDS $3.6\mu$m.

We finally convert the TFIT output into $\mu$Jy using the low-resolution image zeropoints.

\section{Combining the final multiwavelength catalog}

The final catalog is built by combining the SExtractor {\it HST} and TFIT ground-based/{\it Spitzer} 
catalogs. It contains $35932$ sources ($34912$ with a flag $= 0$) and $46$ columns. The catalog 
header is provided in Appendix B.


$\bullet$ Sources that fall on bright star spikes and halos in the Subaru data or the UKIDSS $JHK$ data 
are assigned a flux density and uncertainty of $-99$. 

$\bullet$ The UKIDSS data suffer from cross-talk which echoes all objects in the adjacent amplifiers
of each chip. In the $J$-band image, the crosstalk is clearly visible against the background for sources 
brighter than $J < 16.5$. The crosstalk replica on the adjacent channel was shown to have intensity up to 
1\% of the flux of the original source that drops to about 0.05\% beyond the third amplifier \citep[e.g.,~][]{Dye2006}.
However, crosstalk produced by fainter sources remains unidentified and could affect the photometry 
of some sources. Almaini et al. (in prep.) provide further analysis of crosstalk in the UKIDSS data. The 
crosstalk manifests itself as a spurious `image' that can have different profiles (e.g.,~a ring-like shape 
or a positive-negative source, depending if the original object is saturated or not) but often presents 
a strong negative component. In the present catalog, the regions that were strongly affected by crosstalk were 
identified on the bluer band ($J$) --- where the crosstalk is more prominent --- by detecting its negative 
component using SExtractor on the inverted image (then cleaned by eye to avoid abusive flagging). 
The most strongly-affected $98$ sources are assigned a flux density and uncertainty in the 
$J$, $H$ and $K$ bands of $-99$. 

$\bullet$ The $3.6$ and $4.5\mu$m photometry listed in the present catalog was derived from the
SEDS mosaics. Photometry was also derived from the shallower SpUDS-only $3.6$ and $4.5\mu$m 
data for checks but not included to avoid redundancy. 

$\bullet$ We cross-match the source catalog with $210$ (non proprietary) spectroscopic redshifts in the 
CANDELS UDS field (see column \#45). We also indicate the origin (article published or in preparation) 
of the spectroscopic redshift as well as the nature of the source when available (see column \#46). 
The abbreviations used in column \#46 are detailed in Appendix B.
 
$\bullet$ \citet{Simpson2006} presented a catalog of $505$ radio-sources from radio imaging of the SXDS 
with the Very Large Array (VLA). $38$ sources fall within the CANDELS UDS field. We cross-match the VLA
catalog with the $F160W$-selected catalog, using a matching radius of $0.5$~arcsec, and found that $36$ 
VLA sources have a clear $F160W$ counterpart. Likewise, \citet{Ueda2008} listed $1245$ X-ray detected 
sources ($1213$ point sources and $32$ extended source candidates) in the SXDS field. $32$ point sources 
(and $2$ extended sources) fall within the CANDELS field of view with $27$ of them having an unambiguous 
$F160W$ counterpart (matching radius of $1.5$~arcsec). We specify the radio and/or X-ray nature for these 
sources in column \#46. 

\section {Validation tests on photometry}

\subsection{Consistency checks for similar filters}

\begin{figure}
\begin{center}
\includegraphics[width=8.5cm]{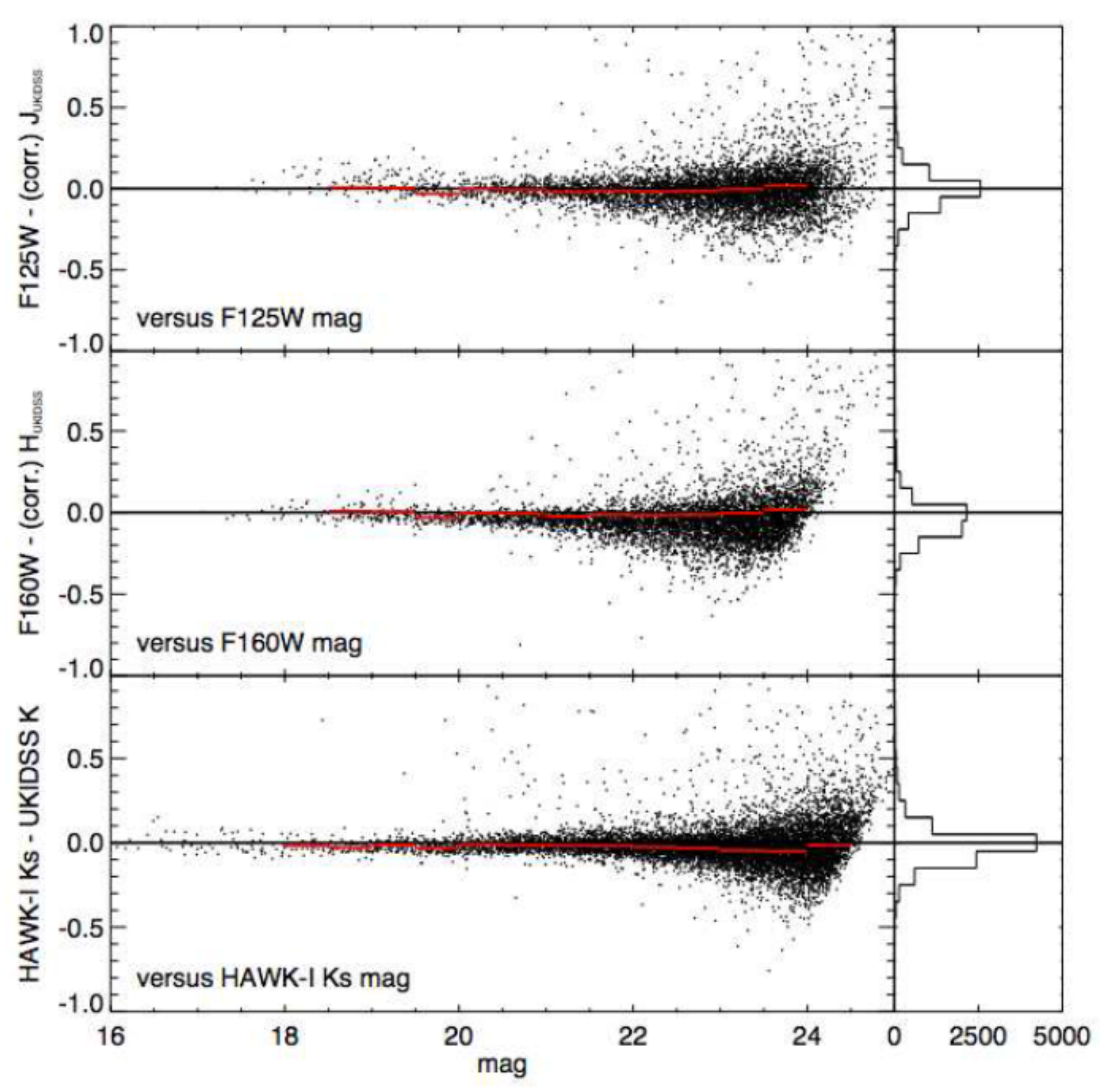}
\end{center}
\caption{Comparison of photometry for similar bands: {\it Top:} between SExtractor $F125W$ and 
TFIT UKIDSS $J$ (versus $F125W$). {\it Middle:} between SExtractor $F160W$ and TFIT UKIDSS $H$
(versus $F160W$). {\it Bottom:} between TFIT HAWK-I $K_s$ and TFIT UKIDSS $K$ (versus HAWK-I $K_s$).
Color corrections were applied in the two upper panels to take into account the filter differences between 
WFC3 and UKIRT. We only plot sources with S/N $> 10$. The histograms of the distributions are
shown in the right insets. The red lines show the $\sigma$-clipped mean per bin of $0.5$mag.}
\label{compJHK}
\end{figure}

$\bullet$ $J$ and $H$ bands: Figure~\ref{compJHK} shows the comparison between the HST/WFC3 ($F160W$, 
and $F125W$) and UKIDSS DR8 ($J$ and $H$) photometry. The WFC3 and UKIDSS WFCAM filters 
are slightly different (see Figure~\ref{trans}). In order to account for the discrepancy between filters, the UKIDSS 
flux densities are converted to the WFC3 photometric system using the color corrections adopted by 
\citet{Koekemoer2011}: $F125W = J + 0.05 (J - H)$ and $F160W = H + 0.25 (J - H)$. In Figure~\ref{compJHK}, 
we only plot sources with $S/N > 10$ and $F160W$ CLASS\_STAR$ < 0.98$ to avoid bright (possibly saturated) 
stars. The agreement is good and shows no systematic offset (i.e.,~no zeropoint issue) or trends.

$\bullet$ $K$ bands: Similarly, we compare the HAWK-I $K_s$ and UKIDSS $K$ photometry (see Figure~\ref{compJHK}). 
No correction is applied because the filters are consistent and correction would be negligible. We also 
find a good agreement between the two bands with no specific discrepancy.

\subsection{Validation tests on colors}

\begin{figure}
\begin{center}
\includegraphics[width=9cm]{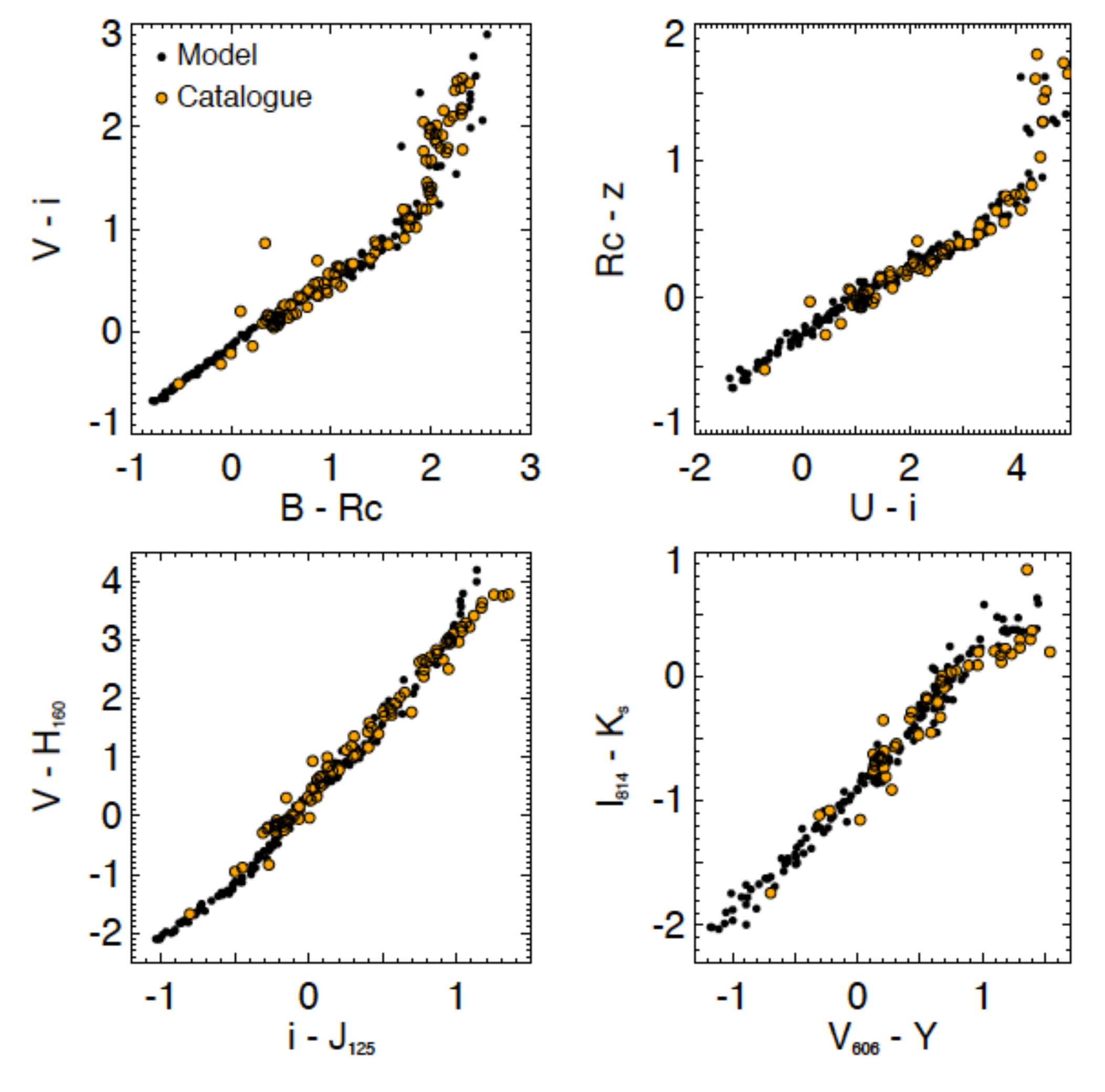}
\end{center}
\caption{Color-color diagrams for synthetic models of stars (black dots) and sources from the catalog 
with CLASS\_STAR $> 0.98$ (orange dots): clockwise from top left: ($V - i'$) vs ($B - R_c$),
($R_c - z$) vs ($U - i$), ($I_{814} - K_{s}$) vs ($V_{606} - Y$) and ($V - H_{160}$) vs
($i - J_{125}$).}
\label{syntstar}
\end{figure}

We study the (optical to near-infrared) colors of stars in the catalog. We first isolate stars in the multiwavelength 
catalog by selecting sources that have CLASS\_STAR $\geq 0.98$. We build a library of synthetic models 
of stars from the Bruzual-Persson-Gunn-Stryker Atlas of stars (Gunn \& Stryker 1983)\nocite{Gunn1983} that we 
convolve with the response curves of the different filters. We then compare their colors to the colors of the stars in 
the catalog and derive a series of color-color diagrams; figure~\ref{syntstar} shows four of these diagrams. The 
agreement is excellent. 

\begin{figure}
\begin{center}
\includegraphics[width=8cm]{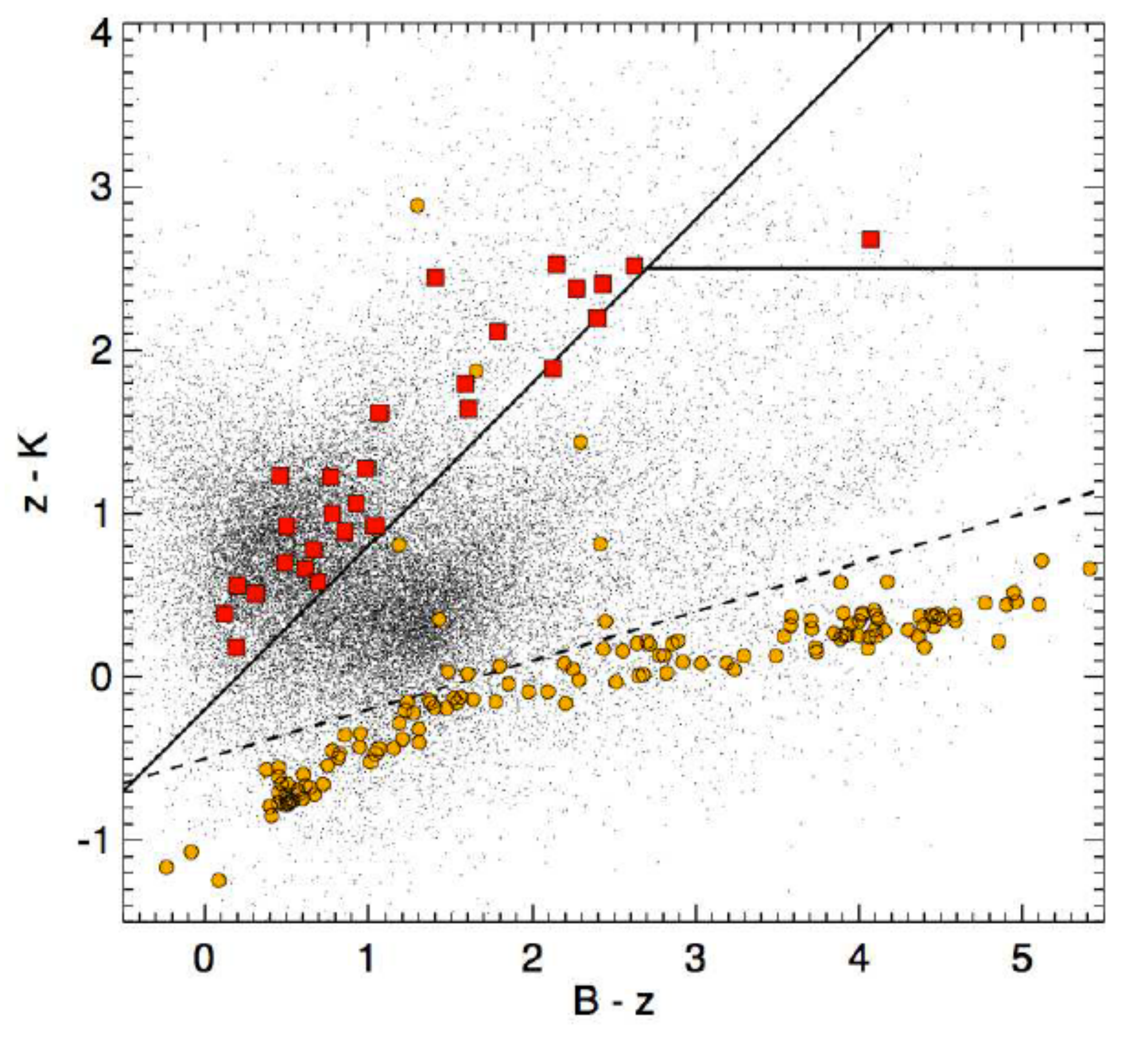}
\end{center}
\caption{$BzK$ color-color diagram of the UDS catalog for sources with S/N$ > 3$ in $B$, $z$ and $K$. 
Color corrections were applied to the $z - K$ and $B - z$ colors to account for the differences in filters
between UDS and Daddi et al.~2004. Sources with CLASS\_STAR $> 0.98$ are shown by the orange 
dots and are found preferentially in the Daddi et al.~(2004) locus for stars i.e.~have 
$(z - K) < 0.3\times(B - z) - 0.5$ (below the dashed line). Sources with spectroscopic redshift 
$1.4 < z_{spec} < 2.5$ are shown by the red squares. All but one 
have colors consistent with star-forming galaxies at $1.4 < z < 2.5$ (s$BzK$-selected galaxies i.e.,~lie 
above the non-horizontal solid line). The exception lies in the p$BzK$ selection area i.e~has colors
consistent with passive elliptical galaxies at $1.4 < z < 2.5$.}
\label{bzk}
\end{figure}

We also look at the distribution of the sources in a $BzK$ color-color diagram (introduced by 
Daddi et al.~2004) which preferentially isolates $z > 1.4$ galaxies. Figure~\ref{bzk} shows the $BzK$ 
diagram derived from the catalog (using the $K$-band photometry from the UKIDSS DR8 data). Small
corrections were applied to account for the differences in filters between UDS and 
Daddi et al.~2004\footnote[14]{$(B - z)_{Daddi} = 1.072 \times (B - z)_{UDS} + 0.06$ and 
$(z - K)_{Daddi} = 1.003 \times (z - K_s)_{UDS} + 0.04$}. Sources with CLASS\_STAR $> 0.98$ (orange dots) 
are, as expected, preferentially found in the stellar locus 
of such diagrams \citep{Daddi2004}. All (but one) sources with spectroscopic redshift $1.4 < z_{spec} < 2.5$ 
(red squares) are s$BzK$-selected galaxies (within error bars) i.e.,~have $BzK \geq -0.2$ (where $BzK = (z - K) - (B - z)$) 
and therefore lie in the typical locus of star-forming galaxies at $1.4 < z < 2.5$. The only source not s$BzK$-selected is found 
in the locus of p$BzK$-selected galaxies ($BzK < -0.2$ and $z - K > 2.5$) i.e.,~has colors consistent 
with a passive galaxy at $1.4 < z < 2.5$.

\begin{figure}
\begin{center}
\includegraphics[width=9cm]{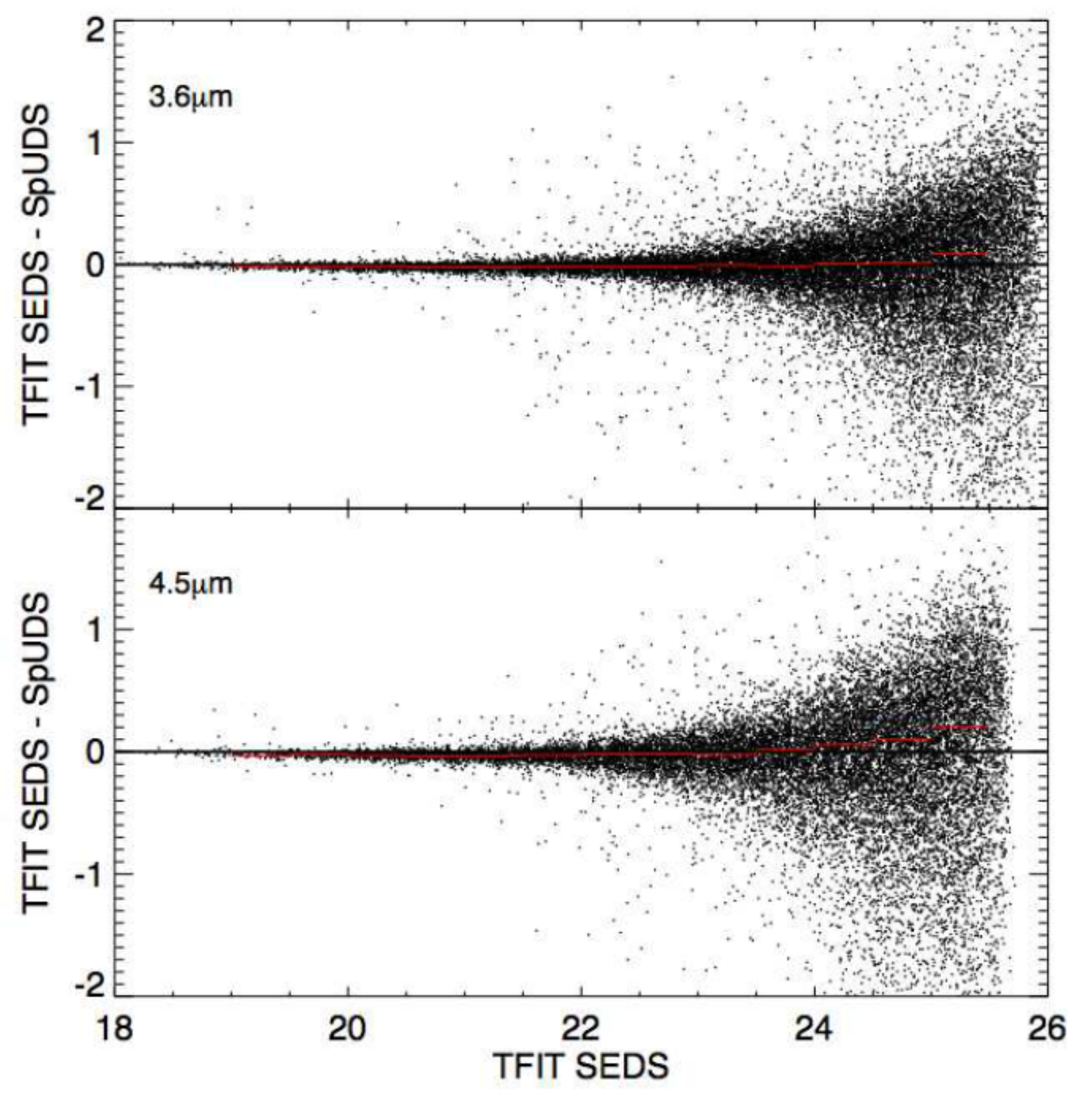}
\includegraphics[width=9cm]{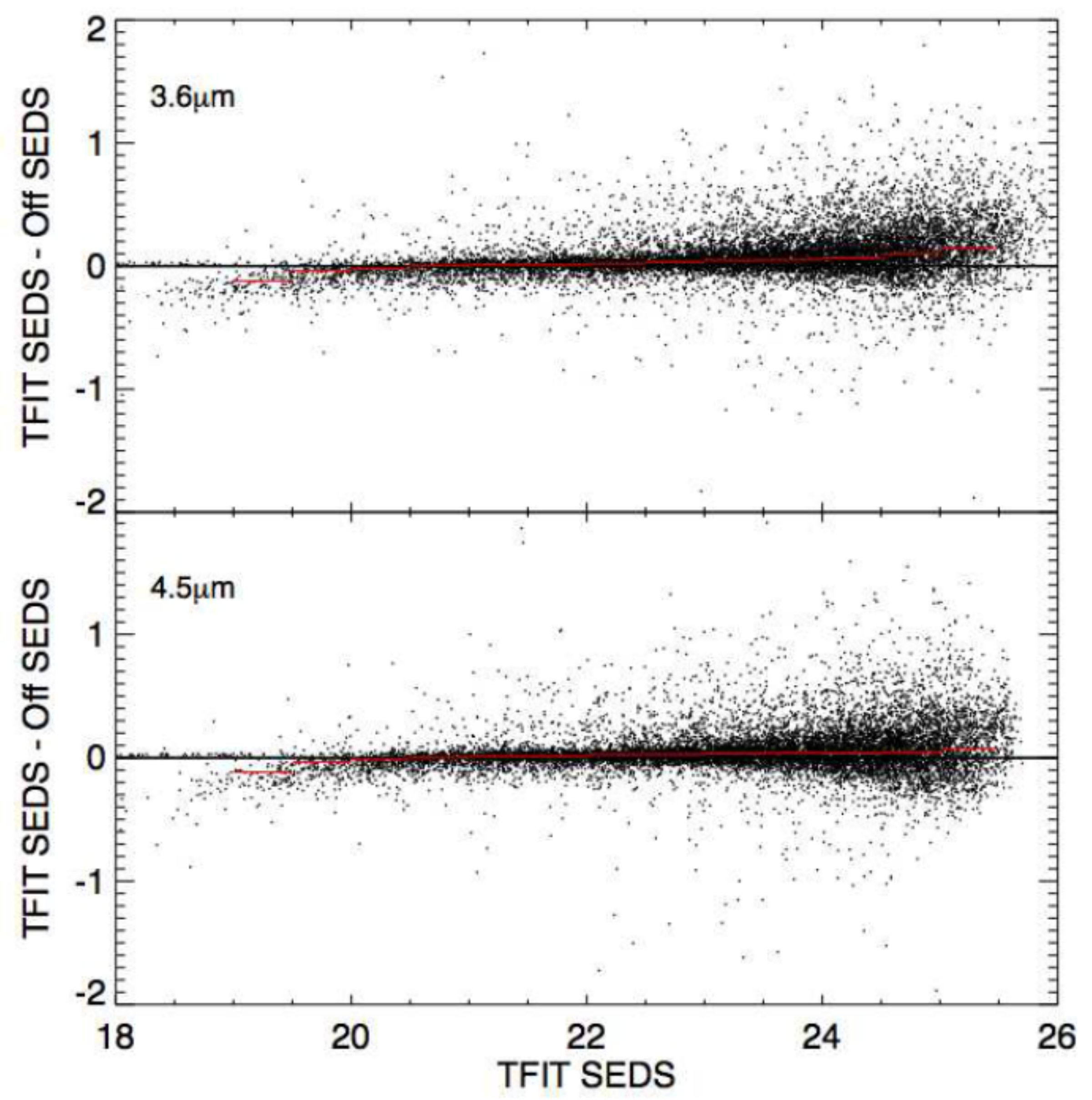}
\end{center}
\caption{Comparison between the TFIT SEDS photometry and TFIT SpUDS-only (top) and the (StarFinder)
SEDS-UDS catalog (bottom) at $3.6$ and $4.5\mu$m for sources with S/N $> 3$. The red lines show the 
$\sigma$-clipped mean per bin of $0.5$mag.}
\label{compirac}
\end{figure}

\begin{figure*}
\begin{center}
\includegraphics[width=12.5cm]{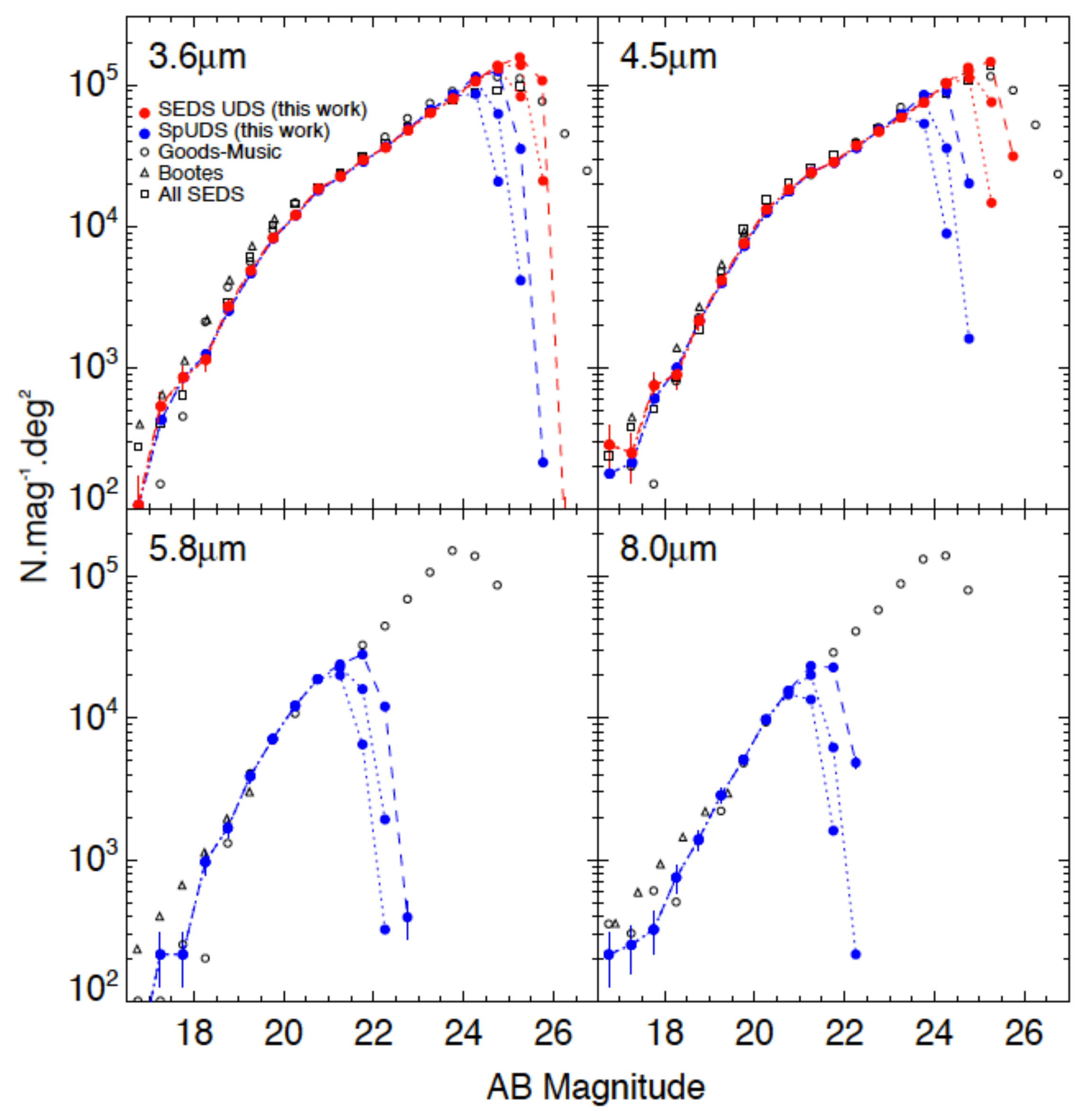}
\end{center}
\caption{Source number counts (stars and galaxies) in the four IRAC bands: $3.6\mu$m (top left), $4.5\mu$m 
(top right), $5.8\mu$m (bottom left) and $8.0\mu$m (bottom right). The counts derived from the present catalog 
(in bin of $0.5$~mag) are shown by the colored dots for SEDS (red) and SpUDS-only (blue). We adopt Poissonnian
errors on the counts. The multiple curves show the counts for different cuts in signal/noise: S/N = $3$ (dashed line) 
and S/N = $4$, $5$ (dotted lines from right to left). Published number counts from \citet{Fazio2004B} (in the Bo\"{o}tes 
field) and Ashby et al.~(resubmitted; for the full SEDS survey) are shown in open symbols (triangles and squares 
respectively). Counts derived from the GOODS-MUSIC are overplotted in open circles.} 
\label{iraccounts}
\end{figure*}


\subsection{{\it Spitzer}/IRAC photometry}

As mentioned earlier, we also derive TFIT photometry for the SpUDS-only $3.6$ and $4.5\mu$m images.
Figure~\ref{compirac} shows the comparison between the TFIT photometry for the SEDS and SpUDS IRAC data. 
The agreement is good with no systematic offset and no suspicious trend with magnitude --- 
i.e.,~the difference is fairly symmetric at a given magnitude --- suggesting that there is no zeropoint offset 
and no systematic background subtraction issue that could bias the faint sources either in the SEDS or 
the SpUDS mosaics.

We also compare our TFIT SEDS photometry to the (StarFinder) SEDS catalog (Ashby et al.~resubmitted).
We made use of an early distribution of their catalog to the CANDELS group (SEDS team; private 
communication) and cross-matched their source list with our $F160W$-selected catalog. In the SEDS 
data, confusion limit is an issue and it is therefore important to use a small aperture in order to avoid flux 
contamination from the wings of nearby sources. We therefore compare the TFIT SEDS photometry to
their aperture-corrected StarFinder photometry derived in a $2.4''$ diameter aperture. The agreement 
is good. The discrepancy increases at faint magnitudes since StarFinder does not deblend faint sources.
Their photometry may be contaminated by neighboring objects and over estimated hence the positive offset.
At the bright end (mag $< 20$), stars have magnitudes consistent in both catalogs. Bright galaxies however
show a larger discrepancy. This was expected when considering magnitudes derived in such a 
small aperture. The match between TFIT SEDS and StarFinder is excellent for bright galaxies 
($< 0.1$~mag) when using the StarFinder $6''$ diameter aperture-corrected photometry. 

Figure~\ref{iraccounts} shows the source number counts derived from the present catalog in the 
four {\it Spitzer}/IRAC bands (SEDS UDS in red and SpUDS-only in blue). Counts are derived for different 
cuts in signal-to-noise (S/N $= 3, 4, 5$) since IRAC photometry at lower S/N becomes rapidly 
unreliable and strongly biased by background contamination issues. We therefore strongly advise 
users to regard the IRAC flux density estimates for sources with S/N $< 3$ with caution. 
Published source number counts from Fazio et al~2004 in the Bo\"{o}tes field and from Ashby et al.~(resubmitted) 
derived from the full SEDS survey (M. Ashby, private communication) are overplotted for information. 
We also derive IRAC number counts from GOODS-MUSIC\footnote[15]{Available at http://lbc.mporzio.astro.it/goods/goods.php}. We precise that the GOODS IRAC $5.8$ and $8.0\mu$m data (shown here as the open 
points) are much deeper than the SpUDS data.

Our IRAC source number counts are in good agreement with previously published number counts. 
As mentioned earlier, the SpUDS data for $3.6$ and $4.5\mu$m are included in the SEDS 
mosaic, although here we measure the photometry independently from a different TFIT run on the 
SpUDS images alone. The number counts are in perfect agreement up to the completeness limit of the SpUDS data
i.e.,~about $1.5$~mag brighter than the SEDS UDS data. The agreement is also excellent with both the 
GOODS-MUSIC and the full SEDS survey (that also includes the present SEDS UDS data). As expected, 
the full-SEDS number counts (derived from the SEDS StarFinder aperture photometry catalog) fall 
below our TFIT SEDS UDS counts (starting $0.5$~mag fainter) since we were able to detect and 
deblend more efficiently the faintest sources thanks to our reference $F160W$-selected source catalog 
coupled with TFIT photometry. 

\begin{figure*}
\begin{center}
\includegraphics[width=12.5cm]{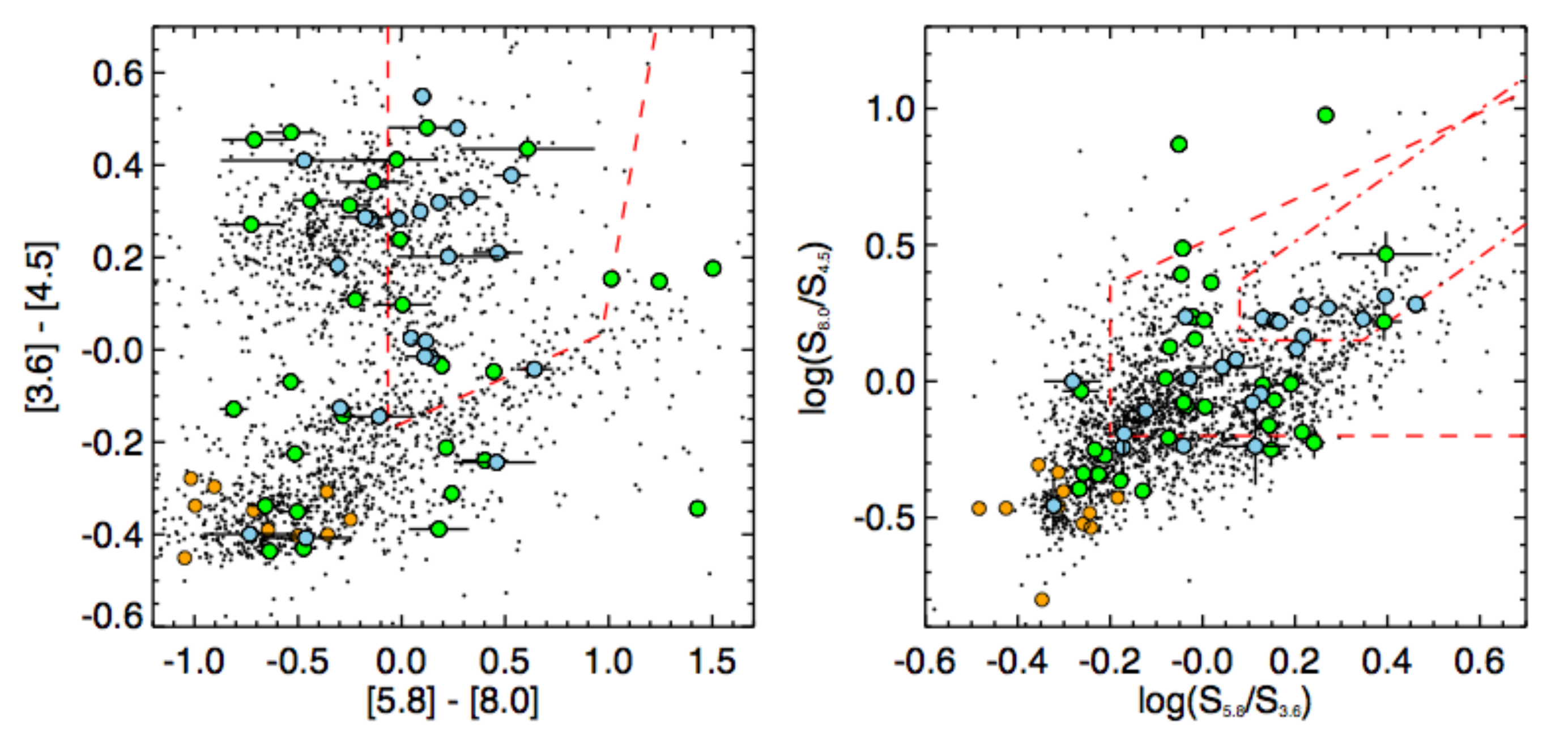}
\end{center}
\caption{IRAC color-color diagrams. {\it Left:} $[3.6] - [4.5]$ versus $[5.8] - [8.0]$ color-color diagram and the
corresponding AGN selection wedge (red dashed line) from \citet{Stern2005}. {\it Right:} log(S$_{8.0}/$S$_{4.5}$) 
versus log($S_{5.8}/$S$_{3.6}$) color-color diagram and its corresponding AGN selection wedge (red dashed line) 
from \citet{Lacy2004}. The 
revised AGN selection criteria from \citet{Donley2012} is also plotted in red dash-dotted line. $3.6\mu$m and $4.5\mu$m 
photometry is from SEDS and $5.8\mu$m and $8.0\mu$m photometry is from SpUDS. We only plot sources 
with S/N $> 3$ in all four IRAC bands. Sources with CLASS\_STAR $> 0.98$ are shown by small orange dots, 
X-ray point sources from \citet{Ueda2008} and radio sources from \citet{Simpson2006} by large blue and green 
dots respectively. Error bars are only shown for X-ray and radio sources for clarity.}
\label{agn}
\end{figure*}

Additionally, we look at the distribution of sources in IRAC color-color diagrams known to 
be an efficient tool to isolate AGN. Figure~\ref{agn} shows the well known mid-infrared AGN selection 
wedges in the $[3.6] - [4.5]$ versus $[5.8] - [8.0]$ color-color diagram first introduced by \citet{Stern2005} 
and the log(S$_{8.0}/$S$_{4.5}$) versus log($S_{5.8}/$S$_{3.6}$) color-color diagram first introduced 
by \citet{Lacy2004} and recently revised by \citet{Donley2012}. We plot only sources with S/N $> 3$ in 
all four IRAC bands. As expected, stars (i.e.,~sources with CLASS\_STAR $> 0.98$; orange dots) have 
colors consistent with zero in the Vega system. The `Stern' selection wedge is more appropriate 
for bright sources and may fail quickly at the depth we are probing in this catalog \citep{Donley2012}. 
Similarly, the `Lacy' wedge is greatly contaminated by faint non-AGN sources and one should think to 
restrict the AGN selection to the less contaminated `Donley' selection wedge.

X-ray point sources (blue dots; Ueda et al.~2008) are found to lie preferentially within the AGN 
selection wedges. Indeed, $61$\% are also mid-infrared-selected according to \citet{Stern2005} AGN 
selection ($78$\% for Lacy et al.~2004), a result somehow surprising when comparing to past studies 
such as \citet{Gorjian2008} which find that only $28$\% of the X-ray detected sources in the Bo\"{o}tes 
field were also mid-infrared selected. The investigation of such discrepancy is however beyond the scope 
of this paper. As expected from past studies \citep{Stern2000}, the overlap of mid-infrared selected AGN 
with radio sources is smaller with only $33$\% ($58$\%) of the ($>$100~$\mu$Jy) radio-sources 
(green dots; Simpson et al.~2006) falling within the `Stern' (`Lacy') AGN wedge.

\begin{figure}
\begin{center}
\includegraphics[width=7cm,bb=100 50 500 400]{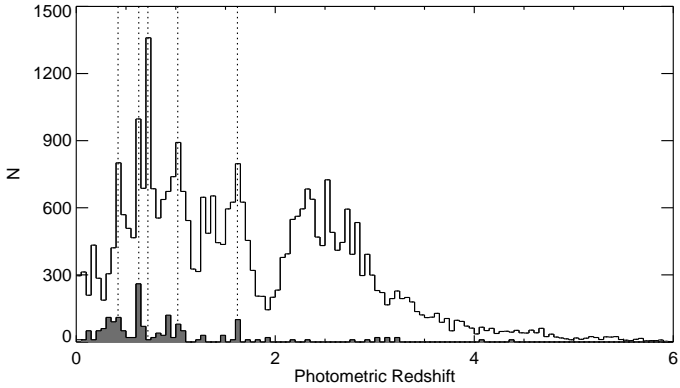}
\end{center}
\caption{Distribution of the photometric redshifts for galaxies (with flag $= 0$) in the CANDELS UDS catalog. 
Several peaks in the redshift distribution e.g.~at $z \sim 0.42$, $0.64$, $0.72$, $1.0$ and $1.62$ (dotted 
vertical lines) confirm the presence of galaxy groups and clusters in the field. The distribution of the spectroscopic 
redshifts available in the UDS catalog is shown by the filled gray histogram (multiplied by a factor $10$ for visibility).}
\label{zpdis}
\end{figure}

\begin{figure}
\begin{center}
\includegraphics[width=8cm]{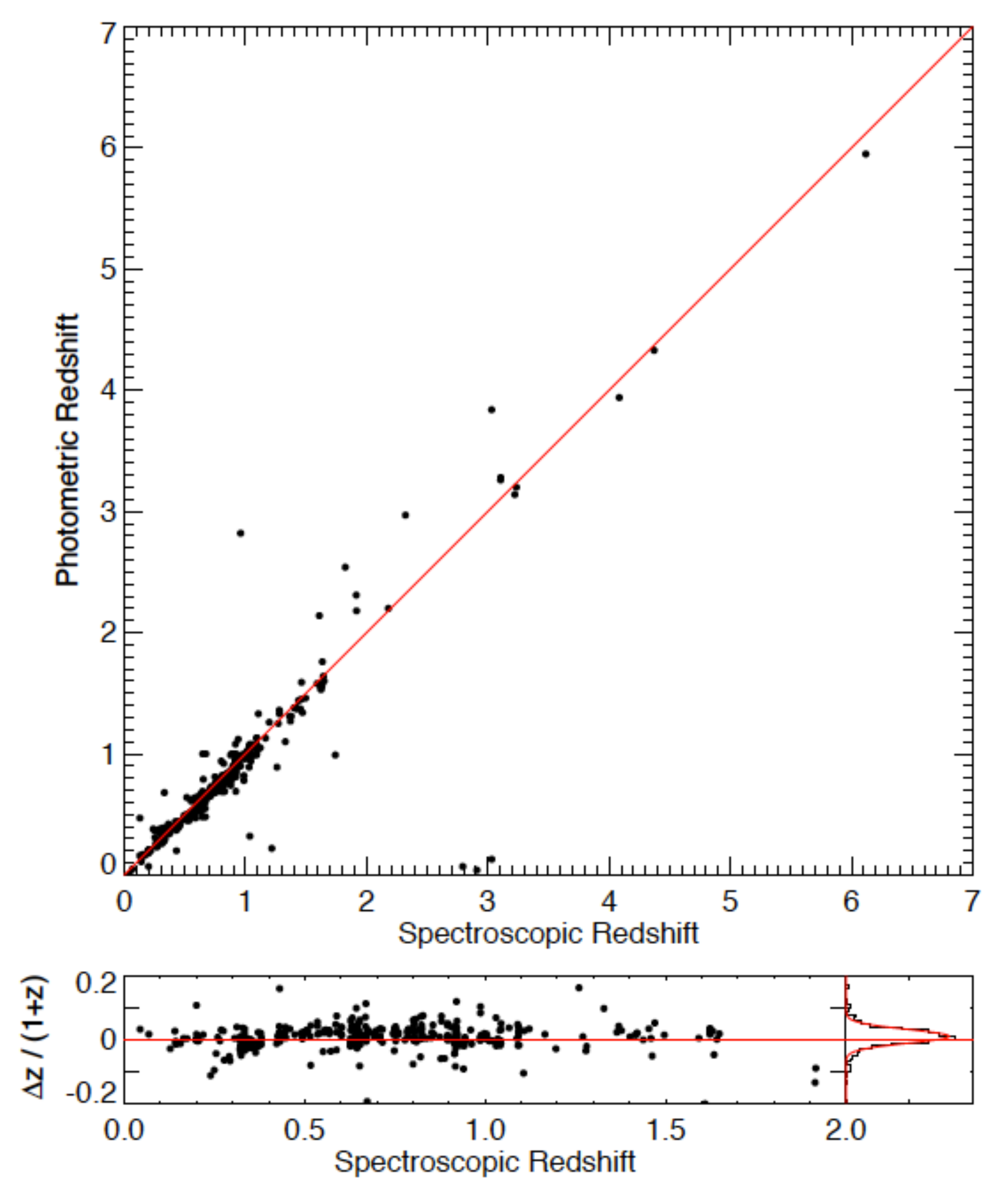}
\end{center}
\caption{{\it Top:} Comparison between photometric and spectroscopic redshifts for galaxies with available
spectroscopic redshift in the UDS catalog. 
{\it Bottom:} Difference $\Delta$$z$ / ($1 + z_{spec}$) where $\Delta$$z = z_{spec} - z_{phot}$ 
restricted for galaxies with $z_{spec} < 2$. The right inset shows the distribution of this difference (solid histogram; 
bin $= 0.01$) with a Gaussian distribution with a standard deviation $\sigma = 0.022$ overlaid (red curve).}
\label{zp}
\end{figure}

\subsection{Photometric redshifts}

Intense work is currently being done within the CANDELS team to derive robust photometric redshifts and stellar mass
estimates for all sources in the CANDELS multiwavelength catalogs. Dahlen et al.~(in prep.) summarize 
the CANDELS team efforts to {\it (i)} compare photometric redshift and mass estimates from different codes ($13$ in total) 
{\it (ii)} assess how well codes manage to recover the redshifts of objects with known spectroscopic redshift and {\it (iii)} 
determine how well codes reproduce stellar masses of simulated galaxies. The final goal is to converge to one 
unique and optimal photometric redshift and stellar mass estimates recipe that is to be adopted for all CANDELS 
multiwavelength catalogs. 

Photometric redshifts and stellar masses derived from the present UDS multiwavelength catalog will be presented
in an upcoming paper. In this section, we make use of one of the $13$ codes presented in Dahlen et al. in 
prep, namely {\tt zphot} \citep{Giallongo1998} as a first test of the photometry of the present catalog. {\tt zphot} is 
a $\chi^2$-minimization procedure that finds the best-fitting template to the observed colors of a source out of a 
spectral library of galaxies. The spectral library was built using PEGASE 2.0 models \citep{Fioc1997}. We 
refer to Grazian et al.~(2006) and references therein for further details on {\tt zphot}.

We first run {\tt zphot} on sources with available spectroscopic redshifts, derive `flux corrections' i.e.~small 
shifts to be applied to the photometry in order to better match the galaxy templates and then run {\tt zphot} on 
the full catalog.

Figure~\ref{zpdis} shows the distribution of photometric redshifts for all sources with flag $= 0$ in the 
UDS catalog. A series of peaks in redshifts is observed, suggesting the CANDELS UDS field contains 
a number of galaxy groups/clusters. We refer to Galametz et al. in prep. for a more detailed study of 
galaxy overdensities in the UDS. We will just note that two structures were already known in the CANDELS 
UDS field: a galaxy cluster at $z \sim 0.65$ \citep[][see also Figure~\ref{psf}]{Geach2007} and one of the
highest redshift galaxy clusters known to date at $z = 1.62$ \citep{Papovich2010, Papovich2012, Tanaka2010}, 
which are both clearly visible in the photometric redshift distribution. 

A comparison between photometric and spectroscopic redshifts is shown in Figure.~\ref{zp}. The
quality of the photometric redshifts is excellent. We further study the distribution of the relative difference 
$z_{spec} - z_{phot}$ / ($1 + z_{spec}$). Its central peak can be cleanly fit by a tight Gaussian with an 
average scatter of $0.022$. However, a large majority of the sources with a spectroscopic redshift in the UDS 
catalog are AGN and the reliability derived from these sources may not reflect the one of the full catalog, 
especially for normal faint galaxies.

\section{Summary}

$\bullet$ The UDS catalog is based on public data including the CANDELS data from HST 
(WFC3 $F125W$ and $F160W$ and 
ACS $F606W$ and $F814W$), $u$-band data from CFHT/Megacam, $B$, $V$, $R_c$, $i'$ and $z'$ bands 
data from Subaru/Suprime-Cam, $Y$ and $K_s$ bands data from VLT/HAWK-I, $J$, $H$ and $K$ bands data 
from UKIDSS (Data Release 8), and {\it Spitzer}/IRAC data ($3.6$, $4.5$ from SEDS, 
$5.8$ and $8.0\mu$m from SpUDS). 

$\bullet$ The UDS catalog is based on a source detection in the CANDELS WFC3 $F160W$ image using 
a slightly modified version of SExtractor. Two detection modes (`cold' and `hot') were used (and then merged) to 
optimally detect and extract all sources ranging from the largest, most extended, and brightest ones to the 
faintest and smallest. SExtractor was also used to derive the photometry in the other HST bands. The final 
catalog contains $35932$ sources over an area of $201.7$ square arcmin.

$\bullet$ Independent catalogs were already publicly available for the Subaru, UKIDSS (although earlier releases) 
and {\it Spitzer}/IRAC data. However, the present catalog not only combines all ultraviolet to mid-infrared bands 
available in the CANDELS UDS field but also takes advantage of the availability of high-resolution and relatively deep
data in the field (i.e.,~the CANDELS HST data). We have indeed used the {\it a-priori} information of the position and 
morphology of sources measured on the $F160W$ image as priors to derive their photometry in the lower-resolution data 
with the TFIT software.

$\bullet$ We cross-matched the catalog with existing catalogs of X-ray sources and radio sources and included
this information in the catalog.

$\bullet$ We also provided, alongside the photometry, a list of sources with spectroscopic redshifts in a first 
attempt to combine all spectra available in the CANDELS UDS field. More upcoming 
spectroscopic campaigns are planned in the next months hopefully improving the current restricted list 
of publicly available spectra.

$\bullet$ A series of convincing tests was done on the photometry to check for its reliability including 
band-to-band comparison, validation with models and a preliminary study of the photometric redshifts
for sources in the catalog.

$\bullet$ The CANDELS UDS multiwavelength catalog is made publicly available on 
the CANDELS website, the MAST archive, via the on-line version of the article, the Centre 
de Donn\'{e}es astronomiques de Strasbourg (CDS) as well as in the Rainbow Database.

\clearpage
\clearpage

\bibliography{apj-jour,biblio}

\clearpage
\clearpage

\begin{appendix}

\section{Appendix A: SExtractor cold and hot detection modes}

\begin{figure}
\begin{center}
\includegraphics[width=10cm]{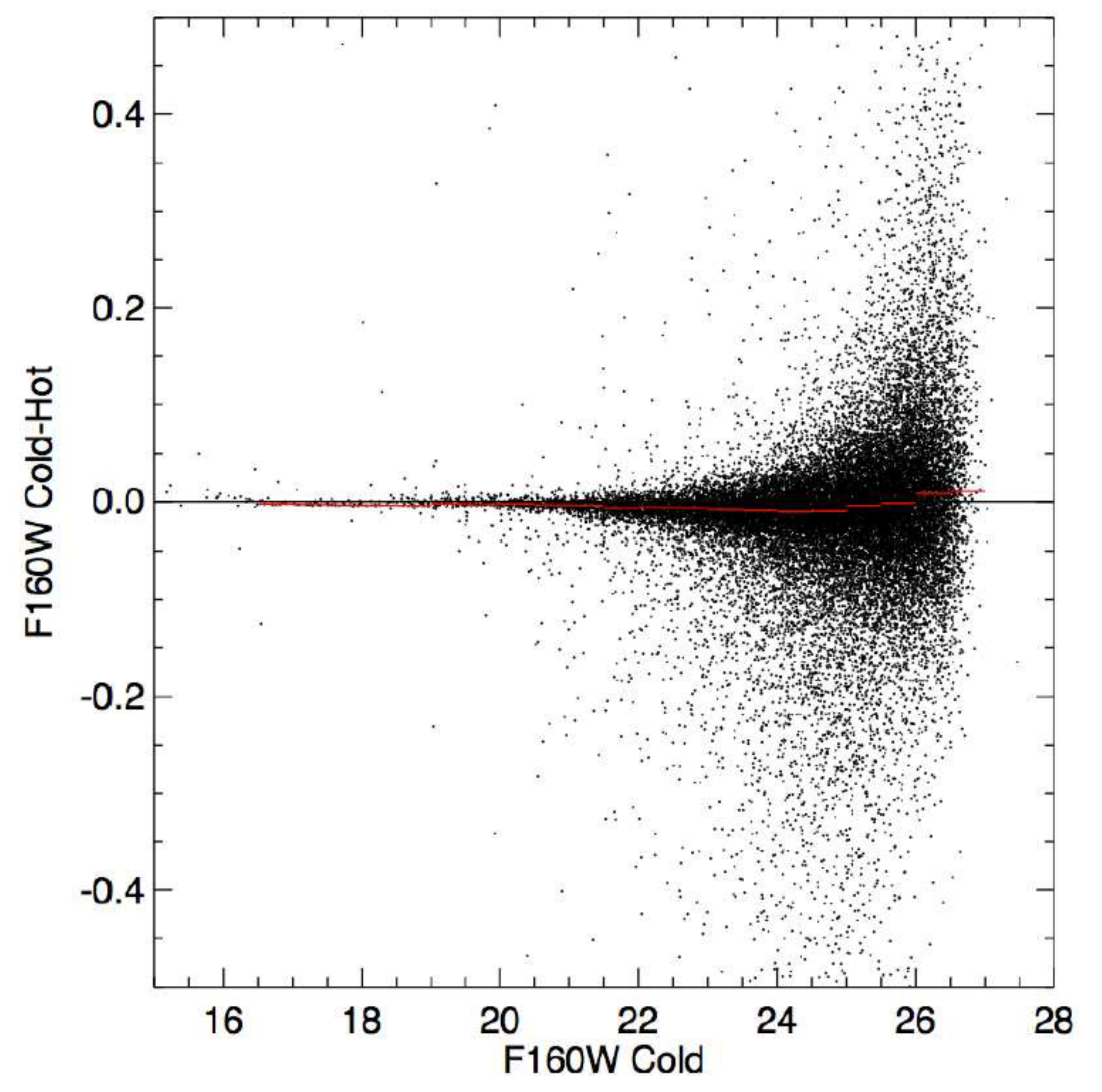}
\end{center}
\caption{Offset in photometry for sources detected in both cold and hot modes (with the 
$\sigma$-clipped mean offset shown in red).}
\label{off}
\end{figure}

The cold mode is optimized for the detection of the brighter and more extended objects. We therefore 
adopt a relatively large smoothing filter (tophat\_9.0\_9x9.conv) to include sub-clumps of large galaxies
within the same galaxy. The hot mode is optimized to detect the fainter sources and we therefore use a 
lower detection threshold, a larger deblending and a smaller smoothing filter. Intensive tests and visual
inspection were made to optimize these parameters to the CANDELS UDS $F160W$ data. We thoroughly
tweaked these parameters to avoid any discontinuity between the cold and the hot mode photometry, 
especially at the faint end when the cold and hot mode start to be merged by the cold + hot routine. Figure~\ref{off} 
shows the agreement in photometry between the two extraction modes for sources detected in both catalogs.\\

{\noindent}\# SExtractor parameter file\\
\# Differences between the two modes are marked in bold following the scheme `Cold / Hot'\\

\#---------- Catalog ---------- \\
CATALOG\_TYPE		ASCII\_HEAD	\\
PARAMETERS\_NAME		file.param \\
CATALOG\_NAME		HST.cat \\
\#---------- Extraction ---------- \\
DETECT\_TYPE		CCD \\
FLAG\_TYPE       		OR \\
{\bf DETECT\_MINAREA	5.0 / 10.0} \\	
{\bf DETECT\_THRESH	0.75 / 0.7} \\
{\bf ANALYSIS\_THRESH	5.0 / 0.7} \\
FILTER				Y \\
{\bf FILTER\_NAME	(cold) tophat\_9.0\_9x9.conv} \\
{\bf FILTER\_NAME	(hot) gauss\_4.0\_7x7.conv} \\
{\bf DEBLEND\_NTHRESH	16 / 64} \\
{\bf DEBLEND\_MINCONT	0.0001 / 0.001} \\
CLEAN				Y \\
CLEAN\_PARAM		1.0 \\
MASK\_TYPE			CORRECT\\
\#---------- Photometry ---------- \\
GAIN 				HST.gain \\
MAG\_ZEROPOINT 		HST.zp \\
PHOT\_FLUXFRAC   	0.2, 0.5, 0.8 \\
PHOT\_APERTURES 	1.47, 2.08, 2.94, 4.17, 5.88, 8.34, 11.79, 16.66, 23.57, 33.34, 47.13 \\
PHOT\_AUTOPARAMS 	2.5, 3.5 \\
{\bf SATUR\_LEVEL     	120.0 / 3900.0} \\
PIXEL\_SCALE     		0.060 \\
MAG\_GAMMA			4.0 \\
\#----------Star/Galaxy Separation ---------- \\
SEEING\_FWHM		0.18 \\
STARNNW\_NAME		default.nnw \\
\#---------- Background ---------- \\
{\bf BACK\_SIZE		256 / 128} \\
{\bf BACK\_FILTERSIZE	9 / 5} \\
BACKPHOTO\_TYPE  	LOCAL \\
{\bf BACKPHOTO\_THICK	100 / 48} \\
\#---------- Weight/Flag Image ---------- \\
\#WEIGHT\_TYPE		MAP\_RMS \\
\#WEIGHT\_IMAGE 		HST.rms.fits \\
WEIGHT\_THRESH   	10000.0, 10000.0 \\
FLAG\_IMAGE			FlagH.fits \\
\#---------- Memory ---------- \\
MEMORY\_OBJSTACK	4000 \\
MEMORY\_PIXSTACK	400000 \\
MEMORY\_BUFSIZE  	5000 \\

\clearpage
\clearpage

\section{Appendix B: Notes on the multiwavelength catalog}

{\noindent}Catalog columns:\\
\# ID (1) 					\\
\# R.A. (deg) (2) 					\\
\# Dec. (deg) (3) 					\\
\# F160W Limiting magnitude (4) 		\\
\# Flag (5) 						\\
\# CLASS\_STAR (6)				\\
\# Flux\_u\_cfht (7)					\\
\# Fluxerr\_u\_cfht (8)				\\	
\# Flux\_B\_subaru (9)				\\
\# Fluxerr\_B\_subaru (10)			\\
\# Flux\_V\_subaru (11)				\\
\# Fluxerr\_V\_subaru (12)			\\
\# Flux\_R\_subaru (13)				\\
\# Fluxerr\_R\_subaru (14)			\\
\# Flux\_i\_subaru (15)				\\
\# Fluxerr\_i\_subaru (16)				\\
\# Flux\_z\_subaru (17)				\\
\# Fluxerr\_z\_subaru (18)				\\
\# Flux\_F606W\_hst (19)				\\
\# Fluxerr\_F606W\_hst (20)			\\
\# Flux\_F814W\_hst (21)				\\
\# Fluxerr\_F814W\_hst (22)			\\
\# Flux\_F125W\_hst (23)				\\
\# Fluxerr\_F125W\_hst (24)			\\
\# Flux\_F160W\_hst (25)				\\
\# Fluxerr\_F160W\_hst (26)			\\
\# Flux\_Y\_hawki (27)				\\
\# Fluxerr\_Y\_hawki (28)				\\
\# Flux\_Ks\_hawki (29)				\\
\# Fluxerr\_Ks\_hawki (30)			\\
\# Flux\_J\_ukidss\_DR8 (31)			\\
\# Fluxerr\_J\_ukidss\_DR8 (32)		\\
\# Flux\_H\_ukidss\_DR8 (33)			\\
\# Fluxerr\_H\_ukidss\_DR8 (34)		\\
\# Flux\_K\_ukidss\_DR8 (35)			\\
\# Fluxerr\_K\_ukidss\_DR8 (36)		\\
\# Flux\_ch1\_seds (37)				\\
\# Fluxerr\_ch1\_seds (38)			\\
\# Flux\_ch2\_seds (39)				\\
\# Fluxerr\_ch2\_seds (40)			\\
\# Flux\_ch3\_spuds (41)				\\
\# Fluxerr\_ch3\_spuds (42)			\\
\# Flux\_ch4\_spuds (43)				\\
\# Fluxerr\_ch4\_spuds (44)			\\
\# Spectroscopic redshift (45)			\\
\# Reference (46)					\\
	
{\noindent}Column description:\\
{\it The electronic version of the table contains some extra columns including 
additional SExtractor parameters derived from the $F160W$ image.}\\
$\bullet$ Column \#1: ID number of the source in the $F160W$-selected SExtractor catalog.\\
$\bullet$ Columns \# 2-3: Right Ascension and declination of the source (J2000) in the $F160W$ image.\\
$\bullet$ Column \# 4: Limiting magnitude at the position of the source in the $F160W$ image (see 
Section 3.3 for details).\\
$\bullet$ Column \# 5: Flag. A specific flag coding is used to designate suspicious sources that fall 
in contaminated regions. A non-contaminated source has a flag of `0'. Sources detected by SExtractor 
at the image edges or on the few artifacts of the $F160W$ image are assigned a flag of `2'. It accounts 
for $230$ sources that are, for a majority of them, not real but that are kept in the catalog to conserve 
the SExtractor original number of sources. Sources detected on star spikes, halos and the bright stars 
that produce those spikes and halos themselves have a flag of `1' ($706$ sources). A large fraction 
of these sources are real but the photometry of these sources --- on which the Template-fitting 
photometry software TFIT run is based (see Section 4) --- is contaminated by a neighbor star.\\
$\bullet$ Column \# 6: CLASS\_STAR parameter in the $F160W$-selected SExtractor catalog.\\
$\bullet$ Columns \#7-44: Flux densities in $\mu$Jy and uncertainties for the $21$ bands of the catalog. 
We consistently report values of $-99$ if the source has no data or is strongly contaminated 
by a star spike  in one specific band.\\
 $\bullet$ Column \#45: Spectroscopic redshift when available; `-99' otherwise.\\
$\bullet$ Column \#46: Origin of the spectroscopic redshift when available; `-99' otherwise. The coding 
follows the scheme `reference-type' (no space):\\

References are coded as follows:\\
`Y05' = Yamada et al. 2005; `G07' = Geach et al. 2007; `Si06' = Simpson et al. 2006; `Si12' = Simpson et al. 2012; 
`Sm08' = Smail et al. 2008; `Ou08' = Ouchi et al. 2008; `V08' = Vardoulaki et al. 2008; `P10' = Papovich et al. 2010; 
`T10' = Tanaka et al. 2010; `F10' = Finoguenov et al. 2010; `SIP' = Simpson et al. in prep.; `AIP' = Akiyama et al. in prep.; 
`CIP' = Cooper et al. in prep.; `PIP' = Pearce et al. in prep.\\

Source types are coded as follows: \\
`NLAGN' = Narrow-line AGN; `BLAGN' = Broad-line AGN; `RadioS' = Radio Source; `RG' = Radio Galaxy; 
`XRay' = X-Ray Source; `QSO' = Quasi Stellar Object; `LAE' = Lyman Alpha Emitter; `ClusterMemb' = Cluster 
member; `OPEG' = Old Passively Evolving Galaxy. \\

Source types for galaxies in the radio source catalog from Simpson et al.~2006 and X-ray source catalog from 
Ueda et al.~2008 are coded as `RadioS(Si06)' and `XRay(U08)' respectively (or both for the only source that 
was detected in radio and X-ray, namely source \#24437). Possible (but questionable) counterparts of X-ray and 
radio sources are indicated by a `?'. Two sources falling within $1$~arcsec of the two X-ray extended source 
candidates (sources \#7217 and \#9461) are coded as `extXRay(U08)'. 

\end{appendix}

\end{document}